\documentclass[]{aa}
\usepackage{graphicx,natbib}
\usepackage{amsmath,amssymb}
\usepackage{mathrsfs}
\usepackage[latin1]{inputenc}
\usepackage[T1]{fontenc}
\bibpunct{(}{)}{;}{a}{}{,}
\usepackage{txfonts}
\usepackage{units}
\usepackage{color}
\usepackage{url}
\newcommand{\ME}{\rm{M}$_{\oplus}$} % w/o $
\newcommand{\RE}{\rm{R}_{\oplus}} 
\newcommand{\Mearth}{\text{M}_{\oplus}} % w/ $
\newcommand{\Msun}{\text{M}_{\odot} }

\newcommand{\Mcore}{M_{\text{core}}}
\newcommand{\Menv}{M_{\text{env}}}

\newcommand{\MHHe}{M_{\text{HHe}}}

\newcommand{\Rp}{R_{\text{P}}}

\newcommand{\Mp}{M_{\text{P}}}
\newcommand{\Miso}{M_{\text{iso}}}

\newcommand{\fhhe}{f_{\text{HHe}} }

\newcommand{\MdotHHe}{\dot{M}_{\text{HHe}}}

\newcommand{\St}{{\rm  S}_{\rm t}}  %Stokes number
\newcommand{\Rh}{R_{\rm H}} %Hill radius
\newcommand{\Hgas}{H_{\rm{gas}}}
 
\newcommand{\SigmaG}{\Sigma_{\rm gas}}
\newcommand{\Md}{M_{\rm d,0}}

\begin{document}

\title{Most super-Earths formed by dry pebble accretion are less massive than 5 Earth masses} 

 \titlerunning{Super-Earths formed by dry pebble accretion}
\authorrunning{Venturini et al.}

\author{Julia Venturini \inst{1}, Octavio Miguel Guilera \inst{2,3,4}, Mar\'{i}a Paula Ronco \inst{3,4}, Christoph Mordasini\inst{5}}
\offprints{J. Venturini}
\institute{International Space Science Institute, Hallerstrasse 6, CH-3012 , Bern, Switzerland.\\
\email{julia.venturini@issibern.ch}
\and
Instituto de Astrof\'{\i}sica de La Plata, CCT La Plata-CONICET-UNLP, Paseo del Bosque S/N (1900), La Plata, Argentina.
\and
Instituto de Astrof\'{\i}sica, Pontificia Universidad Cat\'olica de Chile, Santiago, Chile.
\and
N\'ucleo Milenio Formaci\'on Planetaria - NPF, Chile.
\and
{University of Bern, Gesellschaftstrasse 6, CH-3012, Bern, Switzerland.}}

\abstract
{}
{The goal of this work is to study the formation of rocky planets by dry pebble accretion from self-consistent dust-growth models. In particular, we aim at computing the maximum core mass of a rocky planet that can sustain a thin H-He atmosphere to account for the second peak of the Kepler's size distribution. }
{We simulate planetary growth by pebble accretion inside the ice line. The pebble flux is computed self-consistently from dust growth by solving the advection-diffusion equation for a representative dust size. 
Dust coagulation, drift, fragmentation and sublimation at the water ice line are included.
The disc evolution is computed solving the vertical and radial structure for standard $\alpha$-discs with photoevaporation from the central star. 
The planets grow from a moon-mass embryo by silicate pebble accretion and gas accretion.
We perform a parameter study to analyse the effect of a different initial disc mass, $\alpha$-viscosity, disc metallicity and embryo location. We also test the effect of considering migration vs. an in-situ scenario. Finally, we compute atmospheric mass-loss due to evaporation during 5 Gyr of evolution.}
{We find that inside the ice line, the fragmentation barrier determines the size of pebbles, which leads to different planetary growth patterns for different disc viscosities.
We also find that in this inner disc region, the pebble isolation mass typically decays to values below 5 \ME \, within the first million years of disc evolution, limiting the core masses to that value. After computing atmospheric-mass loss, we find that planets with cores below $\sim$4 \ME \, get their atmospheres completely stripped, and a few 4-5 \ME \, cores retain a thin atmosphere that places them in the gap/second peak of the Kepler size distribution. In addition, a few rare objects that form in extremely low viscosity discs accrete a core of 7 \ME \, and equal envelope mass, which is reduced to 3-5 \ME \, after evaporation. These objects end up with radii of $\sim$6-7 $\RE$.  }
{Overall, we find that rocky planets form only in low-viscosity discs ($\alpha \lesssim 10^{-4}$). When $\alpha \geq 10^{-3}$, rocky objects do not grow beyond Mars-mass. For the successful low viscosity cases, the most typical outcome of dry pebble accretion is terrestrial planets with masses spanning from Mars to $\sim$4 \ME.}


\maketitle

% ----------------------------------------------------- INTRO -----------------------------------------------------------------------------------------------------------------------------
\section{Introduction} \label{intro}
The \emph{Kepler} mission has revealed that most of the stars in the vicinity of the Sun host planets with sizes between that of Earth and Neptune at orbital periods shorter than 100 days \citep{Batalha13, Petigura13}.
Several works have studied the formation of super-Earths/mini-Neptunes, with very different focus. Some addressed the possibility of low-mass cores to accrete thick H-He atmospheres \citep{IkomaHori12, BodLiss14, Lee14, Venturini17}, 
others the formation by giant impacts \citep{Inamdar15, Ogihara18a, MacDonald20,Scora2020, Lee14}, others on understanding the non-resonant configuration \citep{Hands18, Liu2017, Lee_2013}.     

\citet{Fulton17} showed that the Kepler planets present a bi-modal size distribution, with peaks at 1.3 and 2.4 $\RE$.  %, {VanEylen18}. 
The valley in between, at approximately 2 $\RE$ can be reproduced by atmospheric mass-loss mechanisms, such as photoevaporation \citep{Owen17, JinMord18} or core-powered mass-loss \citep{Ginzburg18}. 
Both scenarios are able to reproduce the correct position of the valley only if the naked-cores remaining from the mass-loss are rocky in composition.
This has lead to the interpretation that this population accreted only `dry' condensates and was therefore formed within the water ice line.  
In addition, to reproduce the second peak of the size distribution, mass-loss mechanisms require that the rocky planets start with hydrogen-rich atmospheres that resemble $\sim$1-10\% of the planets' total mass \citep{Owen17}.
From a formation point of view, it is important to evaluate the plausibility to form such rocky objects.

Recently, \citet{Lambrechts19} and \citet{Ogi20} studied the formation of rocky super-Earths by pebble accretion and N-body interactions. 
\citet{Lambrechts19} found  that rocky planets can be produced in two different modes, depending on the pebble flux. 
A total amount of drifted pebbles larger than about 190 \ME \, after 3 Myr would produce `true' super-Earths (planet masses above 5 \ME), while a flux below that threshold would lead to a system of terrestrial planets (planet masses below 5 \ME). The study however neglects gas accretion, so the feasibility of forming 10 \ME \, rocky planets with low mass atmospheres that could explain the second peak of the size distribution remains elusive in that work.

\citet{Ogi20} assumed a similar setup than \citet{Lambrechts19} but they modeled gas accretion, as well as atmospheric mass loss by impacts and evaporation during 50 Myr after disc dissipation.
They also found that super-Earths with cores larger than 5 \ME \, form in systems with a high pebble flux of 100 \ME/My. 
However, they find that those planets tend to accrete too much gas, typically more than 10\% by mass. 
Indeed, after computing mass loss, they find that the surviving H-He atmospheres remain too thick to account for the second peak  of the size distribution. 
The question of whether rocky planets that retain some H-He can explain the second peak of the size distribution is very important, and as \citet{Ogi20}, 
we address it in this work by computing atmospheric evaporation in the post-formation phase.
 
However, before addressing that question, it is equally important to understand, from a formation perspective, what are the typical masses of rocky cores that can be produced by pebble accretion.
We note that \citet{Lambrechts19} and \citet{Ogi20} assume a fixed pebble size or Stokes number, and also take the pebble flux at time zero as a free parameter, which then decays exponentially with time.
Hence, it is not clear if the high and sustained pebble accretion rates they find and that can form cores of $\sim10$ \ME \, can actually occur in nature. 

In this study, we couple pebble accretion simulations with self-consistent dust and gas disc evolution. 
This means that the pebble accretion rates only depend on the initial dust-to-gas ratio of the disc, the disc initial profile, and its initial mass and viscosity.
We take the standard $\alpha$-viscous accretion discs \citep{PT99, Alibert05, Guilera17b} with photoevaporation \citep{Owen2012}.
The model of dust evolution is based on \citet{Birnstiel12} as implemented later in \citet{Drazkowska16, Drazkowska17} and \citet{Guilera20}. 
In our simulations, a moon-mass embryo accretes the pebbles resulting from the dust growth and gas from the disc. 
The methodology is explained in detail in Sect.\ref{sec_method}.

Pebbles are accreted until the protoplanet reaches the so-called pebble isolation mass \citep{LJ14}. At that stage, the perturbation of the protoplanet onto the disc produces a  pressure bump beyond its orbit. 
A pressure bump is a particle trap \citep{HB2003}, so when it forms pebble accretion onto the protoplanet is halted.
The pebble isolation mass depends on the disc properties, mainly on the disc aspect ratio \citep[e.g.,][]{Lambrechts14, Sareh18, Bitsch18}.
The concept of pebble isolation mass is important because it determines the maximum amount of solids or heavy elements that a planet can acquire during the disc phase \citep{Ormel17, Liu19, Bitsch19a}, although giant collisions could occur once the disc dissipates.
However, this is not the only factor limiting the accretion of pebbles. Since the pebbles drift, the disc can also simply run out of them. 
This is another aspect that cannot be assessed when the pebble flux is assumed as a free parameter, instead of being computed from dust growth in a physical finite disc as we do in this work.

Overall, we find that true super-Earths, in the sense of rocky dominated planets with masses larger than 5 \ME, are not a typical outcome of pebble accretion. 
To understand why, we invite you to read this paper.

\section{Method}\label{sec_method}
We use the code {\scriptsize PLANETALP} \citep{Ronco17, Guilera17b, Guilera19, Guilera20} to model the gas and dust disc evolution. To compute the growth of the planets, we coupled {\scriptsize PLANETALP} with the planet formation code developed by \citet{Venturini16}. In the following sections we summarize the main characteristics of the model. 

\subsection{The gaseous disc}
\label{sec:sec2-0}
The gaseous disc evolves in time by viscous accretion and photoevaporation due to central star.
For this, we first solve the disc's vertical structure: hydrostatic equilibrium, energy transport and energy conservation, considering the viscosity and the irradiation from the central star as the energy sources \citep{Guilera17b}. %\citep{PT99, Alibert05, Guilera17b}. 
The averaged gas surface density and viscosity at the disc midplane are used to solve the radial diffusion equation \citep{Pringle1981}: 
\begin{eqnarray}
  \frac{\partial \SigmaG} {\partial t}= \frac{3}{r}\frac{\partial}{\partial r} \left[ r^{1/2} \frac{\partial}{\partial r} \left( \nu \SigmaG r^{1/2}  \right) \right] + \dot{\Sigma}_W (r), 
\label{eq2-sec2-0}
\end{eqnarray}
where $t$ and $r$ are the temporal and radial coordinates, $\SigmaG$ is the gas surface density, and $\nu= \alpha c_s H_{\rm{gas}} $ the kinematic viscosity, given by the dimensionless parameter $\alpha$ \citep{SS73}, the local sound speed ($c_s$) and the disc's scale height ($H_{\rm gas}$). 
$\dot{\Sigma}_W (r)$ is the sink term due to the X-ray photoevaporation by the central star
computed following \citet{Owen2012}, who derived analytical prescriptions from radiation-hydrodynamic models \citep{Owen2010,Owen2011}. 
For full or primordial discs, the fit to the total mass-loss rate as a function of the stellar mass and X-ray luminosity is given by
\begin{equation}
    \dot{M}_w = 6.25\times10^{-9}\left(\dfrac{M_\star}{1M_\odot}\right)^{-0.068}\left(\dfrac{L_\text{X}}{10^{30}\text{erg~}\text{s}^{-1}}\right)^{1.14}M_\odot\,\text{yr}^{-1}.
    \label{eq:O1}
\end{equation}
At some point, a gap is opened in the gas disc at a few au, and the gas inside this region is rapidly accreted by the central star. Once the gas inside the location in which the gap opened is completely drained, the disc becomes a transitional disc (i.e. has an inner hole). Technically, we define this transition once the gas at each radial bin inside the gap is below $\SigmaG = 10^{-5}$ g cm$^{-2}$. The mass-loss rate is then computed as 
\begin{equation}
    \dot{M}_w = 4.8\times10^{-9}\left(\dfrac{M_\star}{1M_\odot}\right)^{-0.148}\left(\dfrac{L_\text{X}}{10^{30}\text{erg~}\text{s}^{-1}}\right)^{1.14}M_\odot\,\text{yr}^{-1},
\label{eq:O2}
\end{equation}
where, following \citet{Preibisch2005}, $L_\text{X}$ is the X-Ray luminosity of the star, as
\begin{equation}
    \text{log}(L_\text{X}[\text{erg~}\text{s}^{-1}]) = 30.37 + 1.44\,\text{log}(M_\star/M_\odot).
\end{equation}
From normalised radial mass-loss profiles (see eq. B2 and B5 of \citet{Owen2012} Appendix B) and with eq.\ref{eq:O1} and \ref{eq:O2} we can compute $\dot{\Sigma}_W (r)$ that verifies $\dot{M}_w = \int 2\pi r \dot{\Sigma}_W (r)dr$.
The disc is resolved from 0.05 to 1000 au using 2000 radial bins logarithmically equally spaced.

Following observations \citep{Andrews10}, the initial gas surface density is taken as:
\begin{eqnarray}
  \SigmaG &=& \SigmaG^0 \left( \frac{r}{r_c} \right)^{-\gamma} e^{-(r/r_{\rm c})^{2-\gamma}}, \label{eq1-sec2-0}
\end{eqnarray}
where $r_{\rm c}$ is the characteristic or cut-off radius and $\SigmaG^0$ is a normalisation parameter that depends on the disc mass. 
We adopt $r_{\rm c}= 39$~au and $\gamma= 0.9$, which are the mean values of the distributions inferred from observations \citep{Andrews10}.
We vary the total disc mass from 0.01 to 0.1 M$_{\odot}$ and the viscosity parameter adopts values of $\alpha = 10^{-5}, 10^{-4}, 10^{-3}$. 

\subsection{Dust evolution}
\label{secDust}

As in \citet{Guilera20}, the evolution of dust along the disc is computed following the approach presented in \citet{Drazkowska16} and \citet{Drazkowska17}, which is based on the results of \citet{Birnstiel11, Birnstiel12}. 
In this model, the maximum size of the dust particles at each radial bin is limited by dust coagulation, radial drift and fragmentation. 
The dust properties change at the water ice line, which is defined at the place where the midplane temperature equals 170 K. The ice line moves inwards as the disc cools.

The maximum particle size at a given time is given by:
\begin{equation}
  r_{\text{d}}^{\text{max}}(t)= \min(r_{\text{d}}^0 \, \exp(t/\tau_{\text{growth}}), \, r_{\text{drift}}^{\text{max}}, \, r_{\text{frag}}^{\text{max}},
  r_{\text{ddf}}^{\text{max}})
  \label{eq1-sec2-1-0}
\end{equation}
where $r_{\text{d}}^0= 1~\mu\text{m}$ is the initial dust size.
We note that at each radial bin and each time step there is a dust/pebble size distribution between 1 micron and $r_{\text{d}}^{\text{max}}$ given by Eq.(\ref{eq1-sec2-1-0}).
$\tau_{\text{growth}}$ is the collisional growth timescale 
\begin{equation}
  \tau_{\text{growth}}= \dfrac{1}{Z\Omega_{\text{k}}}
  \label{eq2-sec2-1-0}
\end{equation}
being $Z= \Sigma_{\text{d}}/\SigmaG$ the dust-to-gas ratio, and $\Omega_{\text{k}}$ the Keplerian frequency. The maximum size of dust particles limited by radial drift is given by 
\begin{equation}
  r_{\text{drift}}^{\text{max}}= f_{\text{d}} \dfrac{2\Sigma_{\text{d}}v_{\text{k}}^2}{\pi \rho_{\text{d}} c_s^2} \left| \dfrac{d\,\ln P_{\text{g}}}{d\,\ln r} \right|^{-1}, 
  \label{eq3-sec2-1-0}
\end{equation}
where $f_{\text{d}}= 0.55$ \citep{Birnstiel12}, $v_{\text{k}}$ is the Keplerian velocity, $P_{\text{g}}$ the gas pressure, and $\rho_{\text{d}}$ is the mean dust density, 
taking values of $\rho_{\text{d}}$ =$3~\text{g}/\text{cm}^3$ and $\rho_{\text{d}}$=$1~\text{g}/\text{cm}^3$ inside and outside the ice line, respectively.
These values assume a pure silicate composition for the grains within the ice line and of a mixture of ices and silicates beyond it \citep{Drazkowska17}. 
The maximum size of dust particles limited by fragmentation is: 

\begin{equation}\label{rmax}
  r_{\text{frag}}^{\text{max}}= f_{\text{f}} \dfrac{2 \SigmaG v_{\text{th}}^2}{3\pi \rho_{\text{d}} \alpha_t c_s^2}, 
\end{equation}
where $f_{\text{f}}= 0.37$ \citep{Birnstiel11}, and we use a fragmentation threshold velocity $v_{\text{th}}$= $1~\text{m}/\text{s}$ for silicate dust and $v_{\text{th}}$=$10~\text{m}/\text{s}$ for icy dust \citep{Drazkowska17}. 
The turbulence strength parameter, $\alpha_t$, affects the impact velocity of dust particles and their settling. In principle, this quantity could differ from the $\alpha$-viscosity. However, as in our disc model the angular momentum transfer is driven by turbulent viscosity, we consider $\alpha_t= \alpha$ throughout this work \citep[as assumed by previous studies, e.g.][]{OrmelKoba2012,Drazkowska17, Guilera20}.

Eq.~(\ref{rmax}) considers that fragmentation is driven by the turbulent velocities. 
However, if the viscosity in the disc midplane becomes very low, fragmentation could be driven by differential drift. The maximum particle size in this case is given by \citep{Birnstiel12}:
\begin{equation}
  r_{\text{ddf}}^{\text{max}}= \dfrac{4 \SigmaG v_{\text{th}}v_{\text{k}}}{c_s^2 \pi \rho_{\text{d}}} \left| \dfrac{d\,\ln P_{\text{g}}}{d\,\ln r} \right|^{-1}. 
  \label{eq5-sec2-1-0}
\end{equation}

The dust is strongly coupled to the gas, reason for which the time evolution of the dust surface density, $\Sigma_{\text{d}}$, is calculated via an advection-diffusion equation, 
\begin{equation}
  \frac{\partial}{\partial t} \left(\Sigma_{\text{d}}\right) + \frac{1}{r} \frac{\partial}{\partial r} \left( r \,\overline{v}_{\text{drift}} \,\Sigma_{\text{d}} \right) - \frac{1}{r} \frac{\partial}{\partial r} \left[ r \,D^{*} \,\SigmaG   \frac{\partial}{\partial r} \left( \frac{\Sigma_{\text{d}}}{\SigmaG} \right) \right] = \dot{\Sigma}_{\text{d}},  
  \label{eqAdvDif}
\end{equation}
where $\dot{\Sigma}_{\text{d}}$ represents the sink term due to planet accretion and pebble sublimation. We consider that when inward drifting pebbles from the outer part of the disc cross the ice line, they lose 50\% of their mass. 
$D^{*}= \nu / ( 1 + \text{St}^2 )$ is the dust diffusivity \citep{YoudinLithwick2007}, and $\text{St}= \pi \rho_{\text{d}} {r}_{\text{d}} / 2 \SigmaG$ the mass weighted mean Stokes number of the dust size distribution. $r_{\text{d}}$ represents the mass weighted mean radius of the dust size distribution, given by
\begin{equation}
  r_{\text{d}}= \dfrac{ \sum_i \epsilon_i r_{\text{d}}^i }{ \sum_i \epsilon_i },  
  \label{eq2-sec2-1}
\end{equation}
being $r_{\text{d}}^i$ the radius of the dust particle of the species $i$, and $\epsilon_i= \rho^{i}_{\text{d}}/\rho_{\text{g}}$ the ratio between the volumetric dust density of the species i and the volumetric gas density \citep[see][]{Guilera20}. We note that because the mean size is mass-weighted, the maximum and mean sizes are very similar, as shown in \citet{Guilera20}, Appendix A. To compute the weighted mean drift velocities of the pebbles population ($\overline{v}_{\text{drift}}$), we follow the same approach as in \citet{Guilera20} considering the reduction in the pebble drift velocities by the dust-to-gas back-reaction due to the increment in the dust-to-gas ratio \citep[see][for details]{Guilera20}.

The initial dust surface density is given by: 
\begin{equation}
    \Sigma_{\text{d}}(r) = \eta_{\text{ice}} Z_{0} \SigmaG(r) 
    \label{eq:eq8-sec2-1}
\end{equation}
where $\eta_{\text{ice}}$ takes into account the sublimation of water-ice and adopts values of $\eta_{\text{ice}}$=1/2 inside the ice line and  $\eta_{\text{ice}}$=1 outside of it \citep{Lodders09}.
We adopt, $Z_{0}= 0.01$ as the initial dust-to-gas ration of the nominal runs.

 \begin{table}
\caption{Disc parameters that we vary and values adopted.} 
\label{tab1}      % is used to refer this table in the text
\centering                                      % used for centering table
\begin{tabular}{l c l }          % centered columns (4 columns)
\hline                      % inserts double horizontal lines
  & Symbol and units & values adopted \\    % table heading
\hline                                   % inserts single horizontal line
 Disc initial mass  & $\Md$ [M$_{\odot}$] & 0.01, 0.03, 0.06, 0.1   \\ %
 Disc viscosity & $\alpha$ & $10^{-3}$,   $10^{-4}$,  $10^{-5}$ \\
 %Disc initial dust-to-gas ratio & Z$_0$ & 0.01, 0.005, 0.03 \\
 Disc initial $\Sigma_{\text{d}}/\SigmaG$ & Z$_0$ & 0.01, 0.005, 0.03 \\
 %Disc characteristic radius & R$_c$ [au]&  39\\ % 
% Disc profile  & $\gamma$ & 1 \\ 
 %Photoevaporation parameter  &  & \\ 
\hline                                             %inserts single line
\end{tabular}
\end{table}

\begin{figure*}
  \centering
  \includegraphics[width=\textwidth]{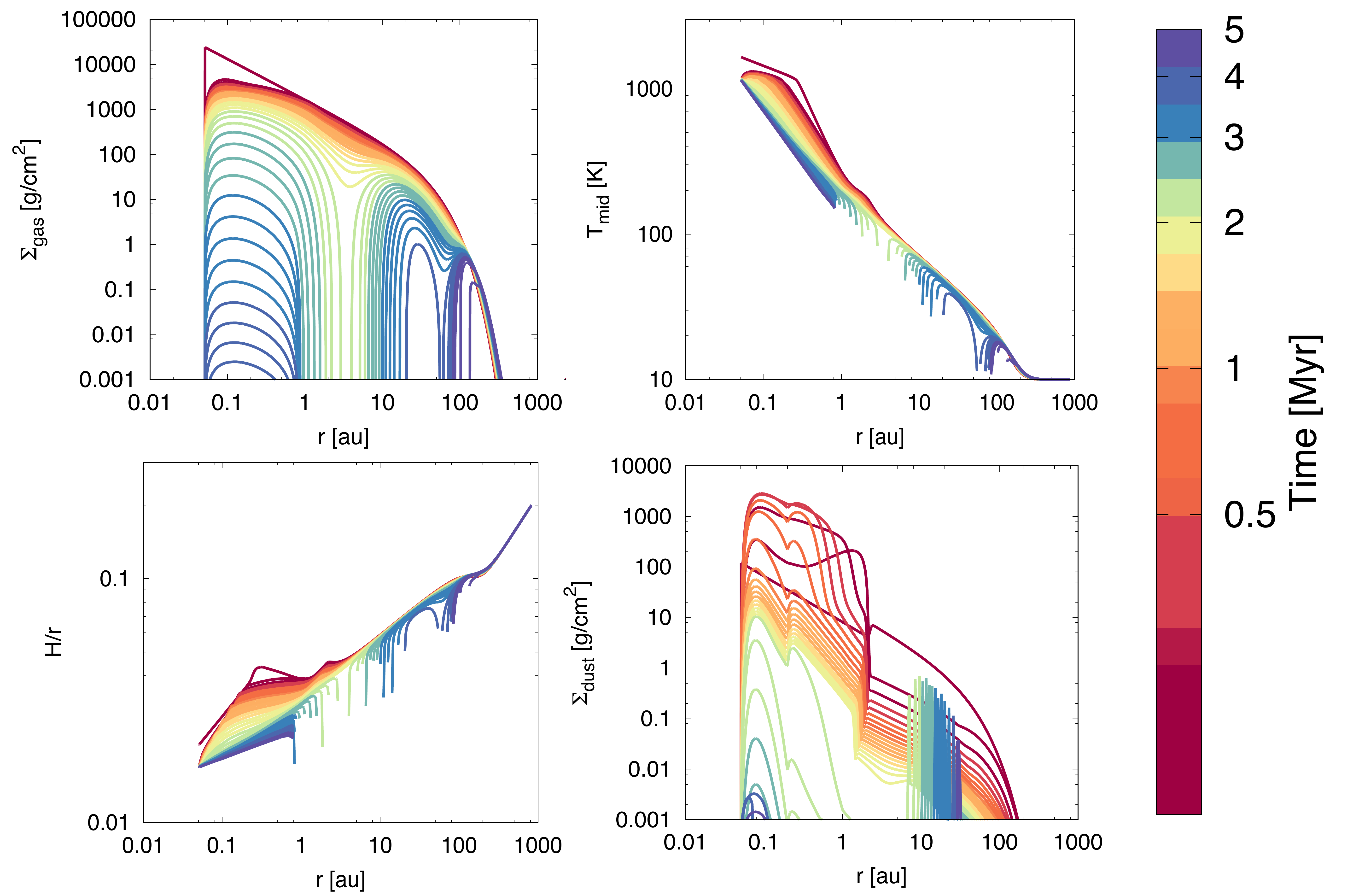}
  \caption{Evolution of the gas surface density (top left), midplane temperature (top right), 
  aspect ratio (bottom left) and dust surface density (bottom right) of the disc with $\alpha=10^{-4}$ and $\Md$ = 0.06 $M_{\odot}$. Profiles every $6\times 10^4$ years. The dip on the dust surface density at 0.2 au is due to the presence of a growing planet at that location.}
  \label{fig_GasDisc}
\end{figure*}

We show an example of gas and dust evolution in Fig.\ref{fig_GasDisc}. The upper and left-lower panels correspond to the evolution of the gas surface density, midplane temperature, and aspect ratio. The right-bottom panel shows the evolution of the surface density of dust. We note that photoevaporation carves a gap in the mid-disc, at time $\sim$2 Myr for this particular case. Once the gap opens, dust from the outer disc cannot reach the inner regions of the system any longer, which leads to a dust accumulation at the outer edge of the gap, visible as verticle lines. Similar dust behaviour in photoevaporating discs were reported in \citet{Ercolano2017}. 

%  FIG 1  

\begin{table*}
\caption{Discs characteristics: gas and dust evolution for the nominal setup ($Z_0 = 0.01$)} 
\label{tab_disc}      % is used to refer this table in the text
\centering                                      % used for centering table
\begin{tabular}{l c c c c c }          % centered columns (4 columns)
\hline                                   % inserts single horizontal line

 $\alpha$ & $\Md$ [M$_{\odot}$] & Time of disc & Time when gap is carved in & M$_{\rm dust}$ after  & M$_{\rm dust}$ after  \\ 
  & &  dissipation [Myr] &  mid-disc by photoevaporation [Myr]  & 0.5 Myr [\ME] & 1 Myr [ \ME]  \\%
 
 \hline                      % inserts double horizontal lines
		 & 0.01 & 0.65 & 0.38 & 3.50 & 0 \\
$10^{-3}$   &0.03 & 1.72 & 1.07 & 8.59 & 1.71 \\
		& 0.06& 3.16 & 1.87 & 25.1 & 5.59 \\
		& 0.1& 4.89 & 2.72 & 53.1 & 12.1 \\
\hline
 		& 0.01 & 2.83 & 0.31 & 6.58 & 2.25 \\
$10^{-4}$  &0.03 & 3.68 & 1.02 &22.9 & 1.64 \\
		& 0.06& 4.90 & 2.15 & 60.7 & 4.87 \\
		& 0.1& 6.70 & 3.64 & 116  & 9.93 \\
\hline
 		& 0.01 & 22.7 & 0.28 & 16.5 & 11.8 \\
 $10^{-5}$  &0.03 & 25.3 & 0.85 & 51.6 &  39.5 \\
		& 0.06& 27.5 & 1.75  & 102 &  78.1 \\
		& 0.1& 29.7 & 3.02 &  164 & 122 \\

\hline                                             %inserts single line
\end{tabular}
\end{table*}

% Pebble Accretion 
\subsection{Pebble accretion}\label{sec:PebbleAcc}
The initial micro-meter size dust grows to mm-cm pebble sizes during the disc evolution.\footnote{We note, however, that formally we do not make any distinction between `dust' and `pebble'. Both refer to all grains fulfilling $\St$<1.} These pebbles can be effectively accreted by embryos present in the disc.
We follow the growth of a Moon-mass embryo ($\Mp$= 0.01 \ME) by pebble accretion. 

To introduce the pebble accretion rate it is first useful to define the pebble scale height \citep[][]{YoudinLithwick2007}:
\begin{equation}
H_{\rm peb}  = H_{\text{gas}} \left(1+  \frac{\text{St}}{\alpha_t}\right)^{-1/2}.
\end{equation}
We note that the pebbles will be more concentrated towards the disc midplane if the disc is rather laminar (low $\alpha$) and/or if the Stokes number is large.
In this case, the pebble accretion occurs in a 2D fashion and for St<0.1 its rate is given by \citep{LJ14}:
\begin{equation}\label{Eq_peb2D}
\dot{M}_{z, 2D} = 2 \, \bigg( \frac{\text{St}}{0.1} \bigg)^{2/3} \Rh v_H {\Sigma}_P \ ,
\end{equation}
where $\Rh$ is the planet's Hill radius, $v_H$ the Keplerian velocity at a distance of the Hill radius from the center of the planet, ${\Sigma}_P$ the surface density of pebbles at the position of the planet,
and $\text{St}$ is the particle's Stokes number at the position of the planet. For 0.1<St<1, Eq.\ref{Eq_peb2D} must be evaluated with St=0.1.

Pebble accretion becomes 3D if:
\begin{equation}\label{Eq_f3D} 
f_{3D} \equiv \frac{1}{2} \sqrt{\frac{\pi}{2}}  \, \left(\frac{\text{St}}{0.1}\right)^{1/3} \frac{\Rh}{H_{\text{peb}}} < 1
\end{equation}

In this case, the accretion of pebbles is given by the slower 3D rate \citep[e.g.][]{Brasser17}:
\begin{equation}\label{Eq_peb3D}
\dot{M}_{z, 3D} = f_{3D}  \, \dot{M}_{z, 2D} 
\end{equation}
 
We note that the lower the Stokes number, the easier it is for f$_{3D}$ to be smaller than 1, and therefore the more likely is that pebble accretion happens in 3D.
For example, for $\text{St} = 0.01$, 2D pebble accretion occurs once the Hill radius exceeds approximately 3.4 pebble's scale height, for $\text{St} = 10^{-4}$  it takes $\Rh$ to be 16 $H_{\text{peb}}$.
Indeed, because of the low Stokes number within the water ice line for $\alpha \geq 10^{-4}$, pebble accretion occurs always in 3D for those cases. 
The same was reported in \citet{Lambrechts19}.

\subsection{The pebble isolation mass}
A planetary embryo can grow by accreting pebbles until its mass is large enough to perturb the disc and create a pressure bump beyond the orbit of the protoplanet.
Hydrodynamical simulations find that the pebble isolation mass can be approximated by \citep[][hereafter L14]{Lambrechts14}:

\begin{equation}\label{MisoEq}
\Miso = 20 \, \left(\frac{H_{\text{gas}}/r}{0.05} \right)^3 \Mearth
\end{equation}

More sophisticated prescriptions were found by \citet[][hereafter B18]{Bitsch18}:
\begin{equation}
\Miso =   25 \left(\frac{H_{\text{gas}}/r}{0.05} \right)^3  \bigg[  0.34 \bigg( \frac{ \rm log(0.001)}{ \rm log(\alpha)} \bigg)^4 + 0.66 \bigg]  \bigg[ 1- \frac{ \frac{\partial {\rm ln} P}{ \partial {\rm ln} r }+ 2.5}{6} \bigg]     \, \Mearth
\end{equation}
and \citet[][hereafter A18]{Sareh18}:
\begin{equation}
\Miso = 25  \left(\frac{H_{\text{gas}}/r}{0.05} \right)^3  \sqrt{82.33 \, \alpha + 0.03} \, \Mearth
\end{equation}

\begin{figure}
\begin{center}
	\includegraphics[width=\columnwidth]{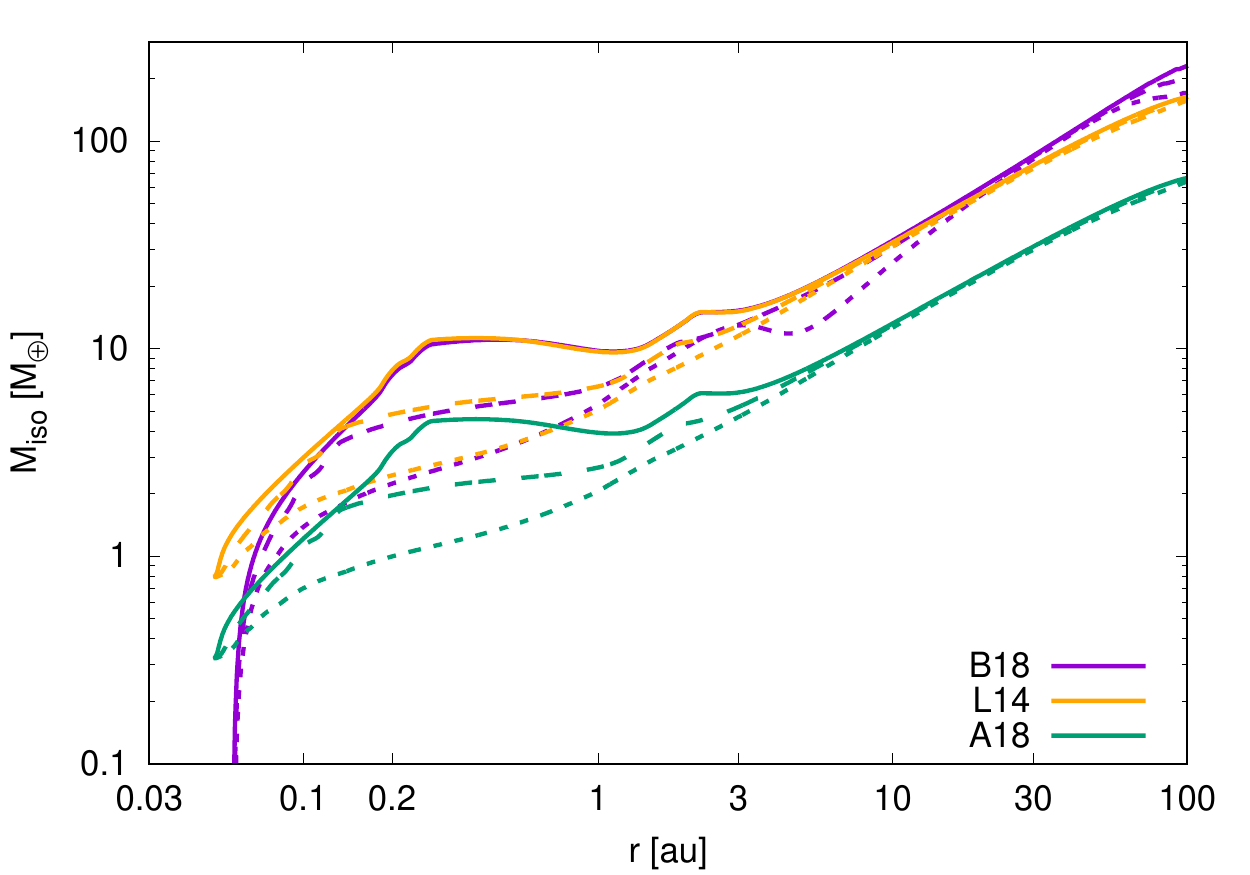}
	\caption{Pebble isolation mass along the disc of $\Md = 0.06 \, \Msun$ and $\alpha = 10^{-4}$ at 20 kyr (solid) , 1 Myr (dashed), and 2 Myr (dotted) of 
	disc evolution.  
	The purple curves correspond to the expression of $\Miso$ found by B18, the green to the one found by A18, and the orange to the one given by L14.
	The inner border of the disc is defined at a=0.05 au.}
\label{fig_miso}
\end{center}
\end{figure}

In Fig.\ref{fig_miso}, we show a comparison among the different definitions of pebble isolation mass for the different prescriptions, for a disc of $\alpha = 10^{-4}$.
We note that the expressions of L14 and B18 give remarkably similar values, while the one of A18 gives lower values. The pebble isolation mass from A18 would only surpass the one of L14 for $\alpha \geq 7.4 \times10^{-3}$, and for such high viscosities the cores do not grow due to the extremely low Stokes number (see Sect.\ref{sec_viscosity}). The same reasoning can be applied for the prescription provided by B18 for $\alpha \gtrsim 10^{-4}$. Since the cores grow only for $\alpha \lesssim 10^{-4}$, and for those viscosities L14 gives the maximum possible $\Miso$, we adopt $\Miso$ from Eq.\ref{MisoEq} throughout this work.

%----------------------------------------------------  INTERNAL STRUCTURE --------------------------------------------------------------------------------------------------
\subsection{Gas accretion}\label{sec_GasAtt}

 In this section we describe the way we compute gas accretion before and after the pebble isolation mass is reached by the protoplanet.

\subsubsection{Attached Phase}\label{sec_attached}
We compute the gas accretion by solving the standard planetary internal structure equations, 
assuming a uniform luminosity that results from the accretion of solids and envelope contraction \citep[see][for details]{Alibert13, Venturini16, Venturini20}.  

The internal structure equations are solved using as boundary conditions the pressure and temperature of the protoplanetary disc at the position of the planetary embryo and defining the planetary radius as a combination of the Hill and Bondi radii, as suggested by 3D hydrodynamical simulations \citep{Lissauer09}: 
\begin{equation}
	R_{\rm P} = \frac{G \Mp}{ \left( c_{\rm s}^2 + 4 G \Mp / \Rh \right) } 
\end{equation}
where $c_{\rm s} $ is the sound speed in the protoplanetary disc at the planet's location $a$,
and the $\Rh$ is the Hill radius. 

We assume that the solids reach the core, meaning that the envelope keeps a H-He composition, and we implement the EOS of \citet{SCVH}.
We define the envelope opacity as $\kappa =  f_{\rm dust} \,\kappa_{\rm gr} +  \kappa_{\rm gas}$ \citep[e.g.][]{Ikoma00}, where the grain opacity is taken from \citet{BL94}, and the gas opacity from the tables of \citet{Freedman14}. 
The grain opacity from \citet{BL94} corresponds to small grain sizes, typical of the ISM. $f_{\rm dust}$ is a reduction  factor to account for the decrease in grain opacity caused by grain growth and settling within the envelope. Studies show that this reduction factor can be typically $\sim$ 0.01 \citep[e.g.][]{Movshovitz10,Mordasini14}.
However, recent work suggest that after attaining $\Miso$, the dust opacity might actually increase due to the trapping of the largest pebbles at the pressure bump \citep{Chen20}.
We are interested in finding the maximum mass of a rocky core to trigger substantial gas accretion, reason for which we stick to conservative high dust opacities and take $f_{\rm dust} = 1$ when solving the internal structure.

After the pebble isolation mass is reached, the resolution of the internal structure equations with uniform luminosity becomes unstable due to the proximity to critical mass. 
Therefore, at this stage we adopt the classical gas accretion rates of \citet{Ikoma00}, which are suited for the case of no solid accretion like ours. 
The prescription is given by the timescale to accrete gas, which depends on the core mass and dust opacity:
\begin{equation}\label{Eq_Ikoma}
\tau_{\rm HHe} = 10^8 \, \bigg(\frac{\Mcore}{\Mearth}\bigg)^{-2.5}   \bigg( \frac{\kappa_{\rm dust}}{ \rm cm^2 g^{-1}}\bigg) \, \, \rm yr 
\end{equation} 
from which the gas accretion rate is computed as $\MdotHHe = \MHHe / \tau_{\rm HHe} $. 
We note that in Eq.\ref{Eq_Ikoma} the dust opacity is constant throughout the envelope, and it takes values of $\kappa_{\rm dust} = f_{\rm dust}$ cm$^2$/g for $T\leq170$ K, and of $\kappa_{\rm dust} = 0.22 \, f_{\rm dust}$ cm$^2$/g for $170 <T <1600$ K (Ikoma, priv. comm.). Hence, since our simulations assume ISM opacities and take place within the water ice line, $\kappa_{\rm dust}= 0.22$. 

We emphasize that the computation of the internal structure equations just after reaching the pebble isolation mass is important, because an increase of approximately one order of magnitude in the envelope mass occurs once the solid accretion is shut down. This happens due to the loss of thermal support originated by the pebble luminosity.
Strictly speaking, the critical mass is achieved when the luminosity due to the envelope contraction reaches a minimum \citep{Ikoma00}. In practical, to guarantee continuity in the gas accretion rate, we switch to Ikoma's prescription when this rate and the one obtained from solving the internal structure are equal.

% ---------------------- Detached Phase -----------------------------------------------------------------------------------------------------

\subsubsection{Detached phase} \label{sec_det}
At certain point the gas accretion required by the protoplanet's thermal structure exceeds the amount that can be supplied by the disc. 
At this stage the protoplanet `detaches' from the disc and continues accreting gas at the rate of the disc viscous accretion:
 \begin{equation}
	 \dot{M}_{\rm gas, disc} = 3 \pi \nu \Sigma_{\rm gas}
\end{equation}

Mathematically, gas accretion is therefore given by:
\begin{equation}
	\dot{M}_{\rm gas} = \text{min} \, \{\dot{M}_{\rm gas, KH}, \dot{M}_{\rm gas, disc}\}
\end{equation} 

Once in the detached phase, gas accretion can be damped even farther if a gap opens in the gas disc. This occurs when \citep{Crida2006}: 
\begin{equation}\label{eqCrida}
\frac{3}{4} \frac{\Hgas}{\Rh} + \frac{50}{q} \bigg( \frac{\nu}{ a^2 \Omega} \bigg) < 1 
\end{equation}
where $q= \Mp / M_{\rm star}$.
When the gap opens, gas accretion reduces even further and is computed following Eqs.36-39 of \citet{Tanigawa07}.

\subsection{Embryo location and migration}
Despite that planet-disc interactions are expected to cause migration, some mechanisms like the presence of pressure bumps can trap protoplanets at a fixed location \citep[e.g.][]{Guilera17}. 
In particular, \citet{Flock19} showed that the silicate ice line can create such a maximum, halting planetary migration at orbital periods of $\sim$10-22 days. 
For our nominal set up we assume in-situ formation at $a=0.2$ au.
This position corresponds to an orbital period of 33 days for a solar-mass star, and is representative of the location where many super-Earths/mini-Neptunes are observed.
We also study in situ formation at $a=0.1$  and $a = 1$ au in Sect.\ref{sec_a01_1}.

In addition,  we study the possibility of migration from just inside the ice line in Sect. \ref{sec_mig}. 
The Type-I migration prescription adopted corresponds to the one derived by \citet{JM17} and \citet{masset2017}, which includes the possibility of outwards migration due to corotation and thermal torques, 
although outward migration does not take place in the cases studied in this work. 
Planets switch to type-II migration once a partial gap opens in the gaseous disc (Eq.\ref{eqCrida}) and migrate according to \citet{IdaLin2004a} or \citet{Mordasini2009} depending on whether they are in the "disc-dominated" or "planet-dominated" type II migration regime \citep{Armitage2007}.

\subsection{Post-formation atmospheric mas-loss }
\label{sec:Christoph}
After the end of the formation phase, we follow the thermodynamic evolution (cooling and contraction) of the individual planets over Gigayear timescales, including the effect of XUV-driven atmospheric photoevaporation. This in particular yields the radii of the planets at 5 Gyr as an important observable quantity. The planet evolution model which was described in details in Mordasini et al. (2012), Jin et al. (2014) and  Mordasini (2020) solves the classical spherically symmetric planetary internal structure equations. The equation of state for the iron-silicate core is the modified polytropic EOS of Seager et al. (2007). The EOS of the gaseous envelope is the new H/He EOS of \citet{Chabrier19}. The outer (atmospheric) boundary condition is given by the semi-gray model described in Jin et al.(2014). The opacities are given by the condensate-free opacity tables of Freedman et al. (2013). In contrast to the formation phase, grain opacities are now neglected, as grains quickly settle into the interior once gas accretion stops (Mordasini 2014). When calculating the luminosity evolution, the gravitational and internal energy of the core and envelope as well as radiogenic heating are included (see Linder et al. 2019). Atmospheric escape rates are found by interpolating in the grid of evaporation rates provided by Kubyshkina et al. (2018). These evaporation rates were obtained with the upper atmosphere hydrodynamic code of Erkaev et al. (2016). The code solves the 1D equations for mass, momentum, and energy conservation and accounts for Ly$\alpha$ cooling, XUV heating, and H$^+_3$ cooling. The stellar XUV luminosity as a function of time is taken from McDonald et al. (2019).

% ================================= RESULTS ====================================

\section{Results} \label{sec_results}
Here we present results considering different disc mass, viscosity, metallicity, and formation locations.

\subsection{The key role of viscosity}\label{sec_viscosity}
\begin{figure*}
\begin{center}
	\includegraphics[width=0.95\textwidth]{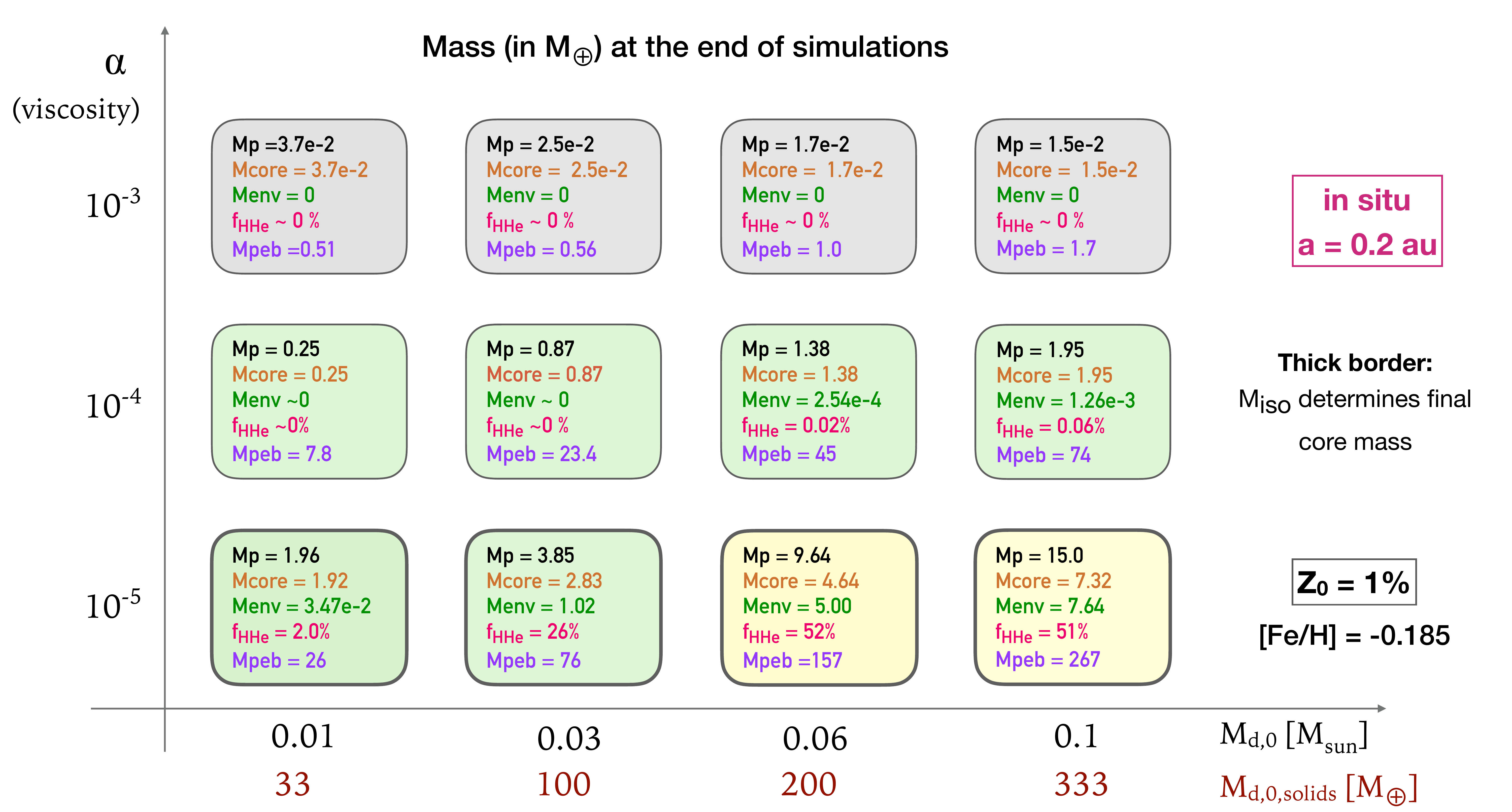}
	\caption{Summary of results for the nominal set-up at the end of the simulations as a function of the initial disc mass and $\alpha$-viscosity. The horizontal axis also shows the initial mass of dust in the disc. $\Mp$ denotes the total planet mass, $\Mcore$ the core mass, $\Menv$ the mass of H-He envelope and $M_{\rm peb}$ the total mass of pebbles that drifted through the embryo's position.  All masses in the squares are in $\Mearth$. $\fhhe$ is the planet's mass fraction of H-He in \%. Cases where planets remain as lunar-mass embryos are colored in grey. 
	Planets with masses larger than Mars and rocky-dominated ($\fhhe$<30\%) are highlighted in green, and gas-dominated planets ($\fhhe$>30\%) in yellow.
	Boxes with thick borders indicate that the core growth was truncated by reaching the pebble isolation mass (see also Table \ref{TabPflux}).}
\label{fig_nominal_res}
\end{center}
\end{figure*}

We start by showing the results of our nominal case, defined as in-situ growth for an embryo located at $a=0.2$ au from the Sun (or orbital period of 33 days). 
The initial disc's dust-to-gas ratio is taken as Z$_0$ = 0.01.
The results for this set-up are summarised in Fig. \ref{fig_nominal_res}.  
We highlighted the cases with different colours: grey shows simulations where the planet practically does not grow and ends up with a mass in the order of magnitude of the moon.
Green remarks rock-dominated planets, defined as having final planet mass larger than Mars and H-He mass fraction of $\fhhe <50 \%$. Finally, we show in yellow gas-dominated planets, which are defined as having $\fhhe \geq 50 \%$. 
We note that the output of the simulations is extremely sensitive to the assumed $\alpha$. In discs with $\alpha =10^{-3}$, the initial moon-mass embryo practically does not grow, for any disc mass. 

\begin{figure}
\begin{center}
	\includegraphics[width=\columnwidth]{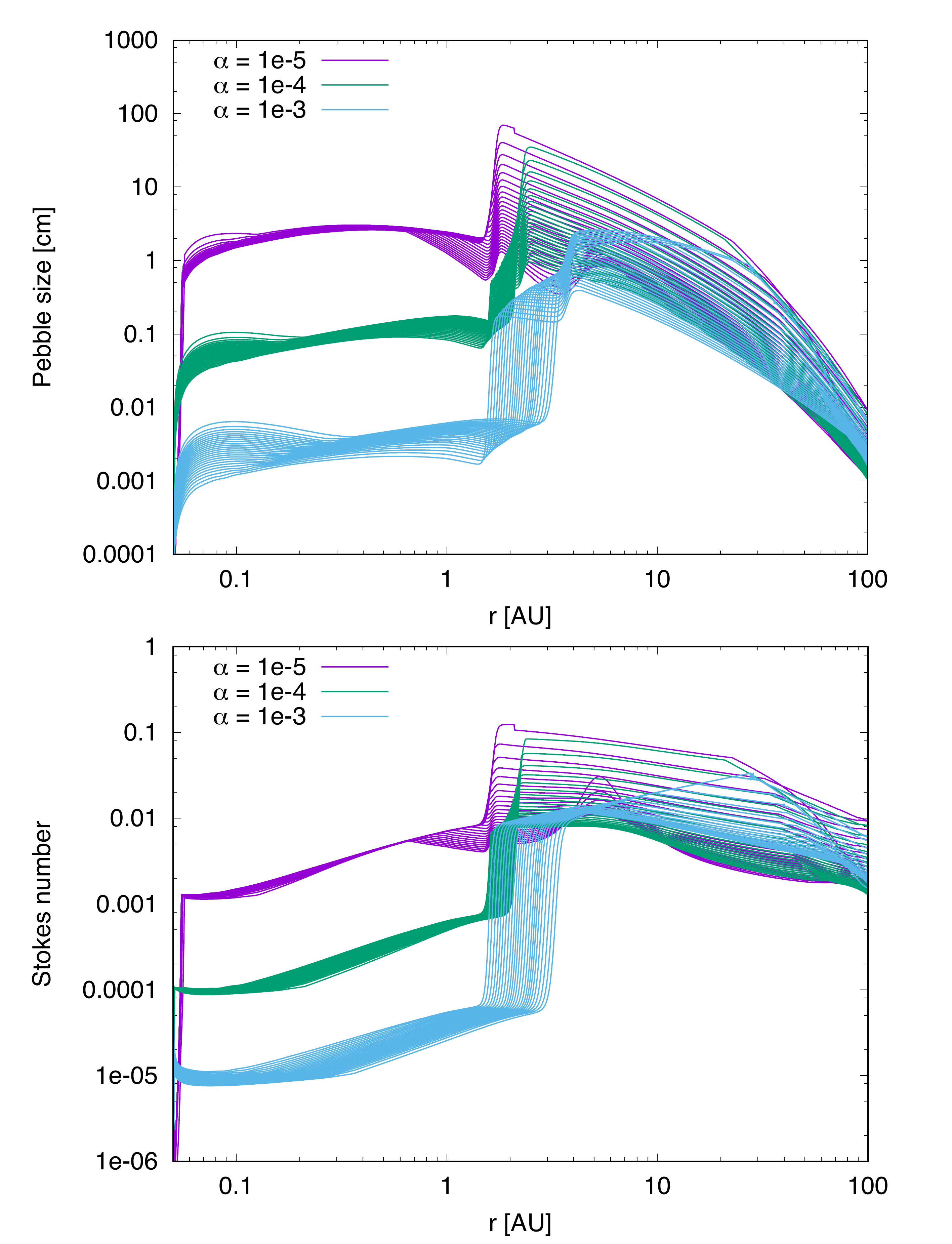}
	\caption{Evolution of maximum pebble (or dust) size (top) and Stokes number (bottom) along the disc of $\Md = 0.06 \, \Msun$. 
	The profiles are shown every  80 Kyr and until photoevaporation carves a gap in the middle of the discs, halting pebble drift. 
	The change of behaviour at $\sim$1-3 au is due to the change of pebbles' properties at the water ice line. }
\label{fig_pebblesize}
\end{center}
\end{figure}

In general, the trend of Fig. \ref{fig_nominal_res} is that the final planet mass increases the lower the viscosity and the larger the initial disc mass. 
Why does this happen?
When we analyse the evolution of the pebble sizes along the disc (Fig.\ref{fig_pebblesize}, upper panel), we see that, inside the  water ice line, the larger the viscosity, the lower the particle size.
Inside the water ice line the fragmentation threshold velocity is low ($v_{\rm th}$ = 1 m/s), implying a maximum particle size limited by fragmentation. 
Turbulence promotes fragmentation (Eq.\ref{rmax}), and therefore, the larger the $\alpha$,  the smaller the pebble size.  
From the definition of the Stokes number, it is clear that the lower the particle size, the lower its Stokes number. 
We observe this in the lower panel of Fig.\ref{fig_pebblesize}. 
Because of fragmentation, the Stokes number decreases by one order of magnitud when $\alpha$ increases by the same amount.
The Stokes number affects the pebble accretion rate in two ways.
First, it enters directly into Eqs.\ref{Eq_peb2D}-\ref{Eq_peb3D}. The lower the Stokes number, the lower the pebble accretion rate. 
Second, it affects the surface density of pebbles, which also enters in Eq.\ref{Eq_peb2D}. 

\begin{figure}
\begin{center}
	\includegraphics[width=0.95\columnwidth]{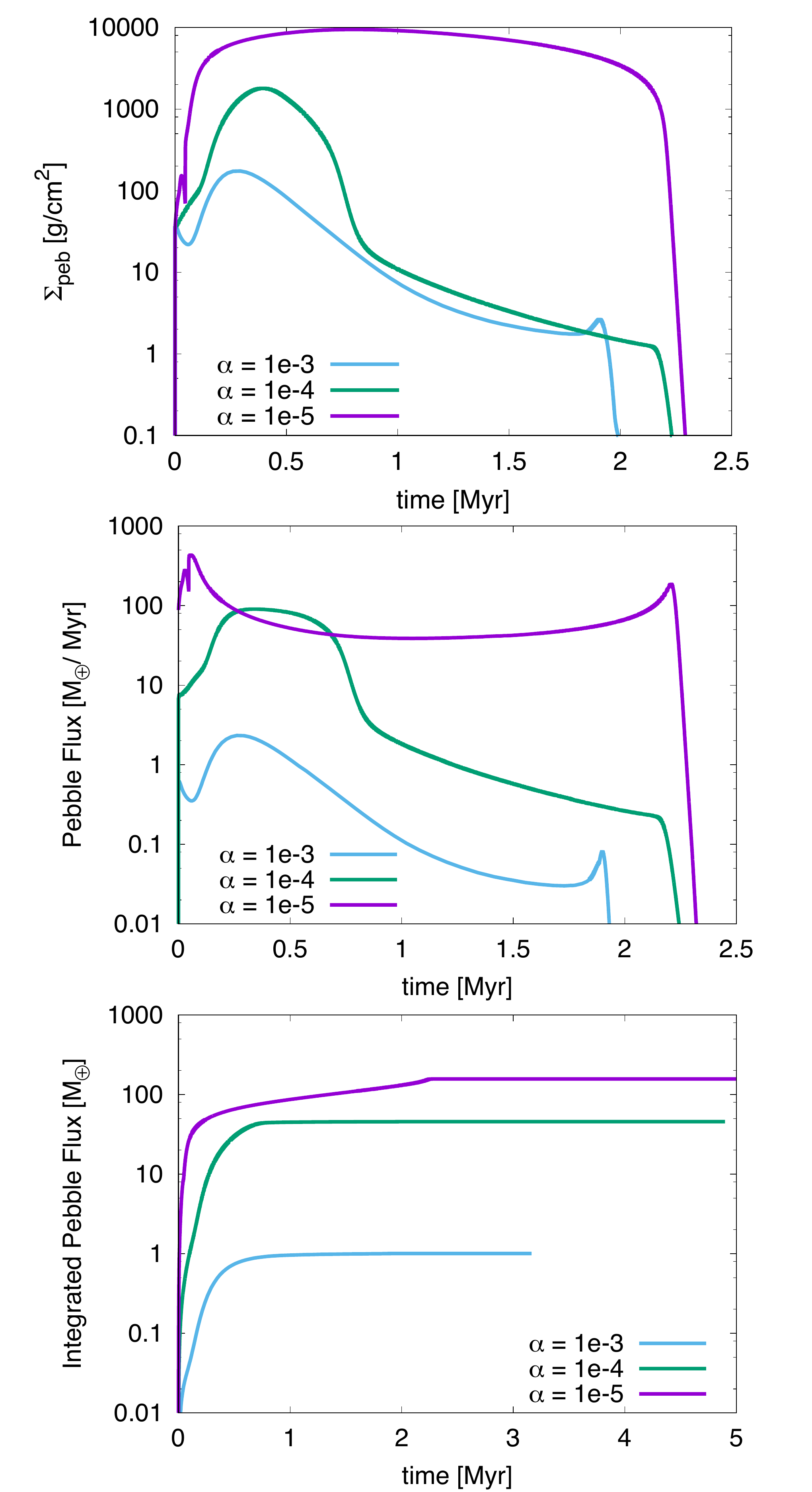}
	\caption{Evolution of the pebble surface density (top), the pebble flux (middle) and the integrated pebble flux (bottom) of the disc with $\Md = 0.06 \, \Msun$ of the nominal set-up (Fig.\ref{fig_nominal_res}).  
	The pebble flux ceases once the gap due to photoevaporation is carved at $a \approx 3-4$ au (see Fig.\ref{fig_GasDisc}).
	}
\label{fig_pebFlux}
\end{center}
\end{figure}

\begin{figure}
\begin{center}
	\includegraphics[width=0.95\columnwidth]{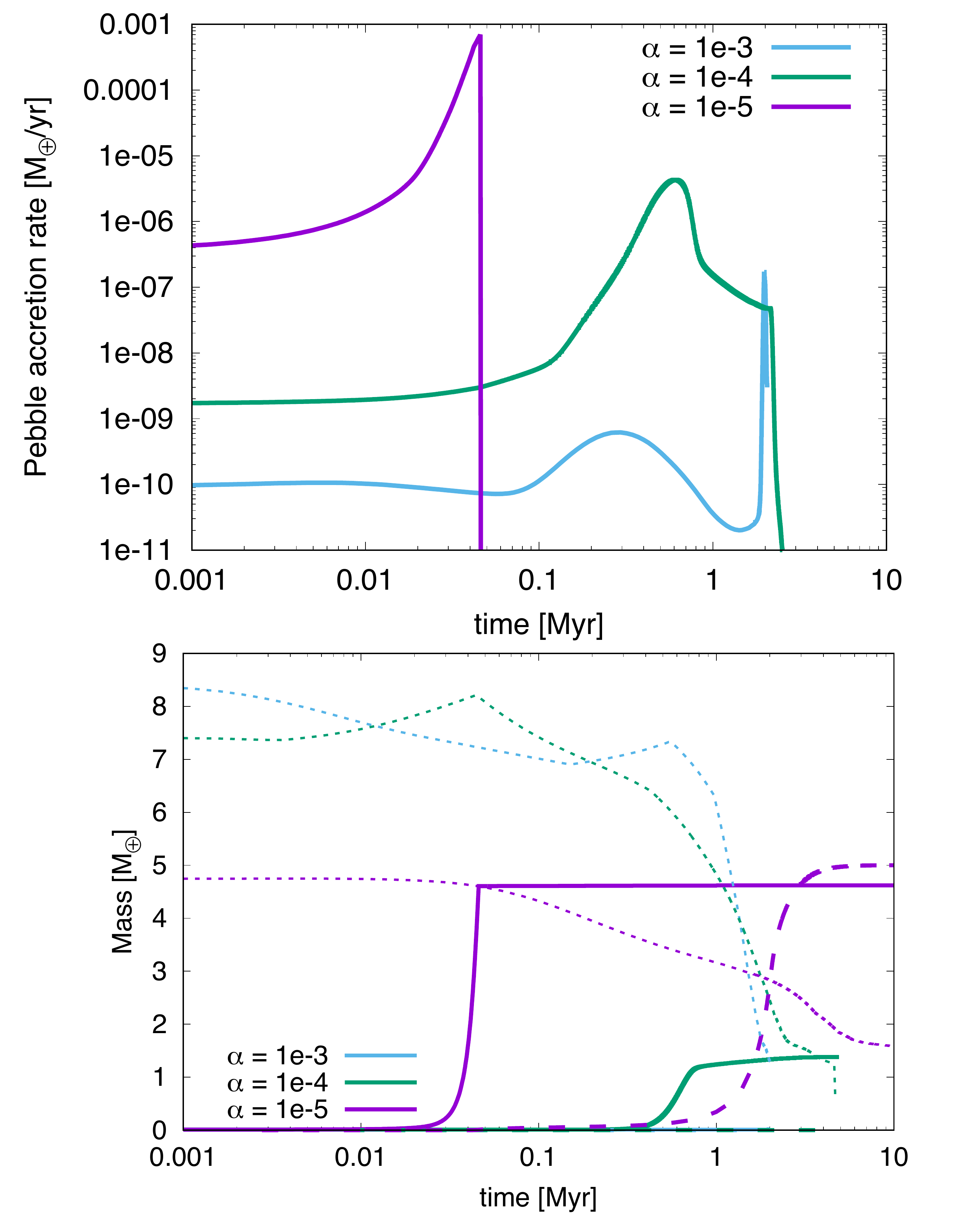}
	\caption{Evolution of the pebble accretion rate (top) and planet growth (bottom) for the same cases as in Fig.\ref{fig_pebFlux} (disc of $\Md = 0.06 \, \Msun$ of the nominal set-up).  
	 In the bottom panel the solid lines shows the growth of the core, the dashed lines the growth of the envelope, and the dotted-thin lines the evolution of the pebble isolation mass for each case. 
	 The accretion rate of pebbles is shut down very early for the case of $\alpha = 10^{-5}$ because $\Miso$ is attained.}
\label{fig_pebAcc}
\end{center}
\end{figure}

The top panel of figure \ref{fig_pebFlux} deploys the surface density of pebbles as a function of time, which results from solving the advection-difussion equation (Eq.\ref{eqAdvDif}) 
We note that both the behaviour of $\Sigma_{\rm peb}$ and stokes number promote core growth for low viscosities.
As long as the Stokes number is lower than 1, the lower its value, the lower the drift velocity (see Eqs. 11-16 in \citet{Guilera20}). 
This means that the pebble flux will decrease with increasing $\alpha$. 
This is shown on the central panel of Fig. \ref{fig_pebFlux}, where we compare pebble fluxes for the different $\alpha$ of the disc with $\Md = 0.06 \, \Msun$.
From this figure it is clear that the pebble flux is not constant, varying up to 2-3 orders of magnitude as the disc evolves.
Importantly, for all cases the pebble flux and the surface density of pebbles decays abruptly at $\sim$ 2-2.5 Myr of disc evolution. 
This occurs because shortly before (see Table \ref{tab_disc}), a gap in the mid disc is carved by photoevaporation (see, e.g. top left and bottom right panels of Fig.\ref{fig_GasDisc}), so the pebble supply from the outer disc is shut down.
The bottom panel of figure \ref{fig_pebFlux} shows the integrated pebble flux for the same cases, in other words, the total amount of pebbles that passed through the position of the planet until a given time.
We note that an integrated pebble flux larger than 190 \ME, needed to form `true' Super-Earths ($\Mp > 5 \Mearth$) in the work of L19, is never achieved. 
This explains why our results favour the formation of terrestrial planets by dry pebble accretion, and not of super-Earths. 
We note as well that after only 1 Myr of disc evolution, the amount of pebbles at 0.2 au has already receded so strongly that any substantial core growth beyond that time is unfeasible, as shows figure \ref{fig_pebAcc}.
The top panel of Fig.\ref{fig_pebAcc} shows the pebble accretion rate onto the protoplanet for the same cases as figure \ref{fig_pebFlux}. These accretion rates occur always in 3D.
We note that the pebble accretion rate varies up to 4 orders of magnitude between $\alpha = 10^{-5}$ and $\alpha = 10^{-3}$. 
This explains the different outputs of the nominal set-up displayed in Fig.~\ref{fig_nominal_res}.
We note in Fig.\ref{fig_pebAcc} that the pebble accretion rate for $\alpha = 10^{-5}$ drops abruptly at $t= 5\times10^{-4}$ yrs, despite that a large pebble flux still exists at that time. 
This happens because the planet reaches the pebble isolation mass at that early time, as can be appreciated better in the bottom panel of the same figure.
For $\alpha = 10^{-4}$ the pebble isolation mass is reached late, at t = 3.7 Myr, when the core basically does not grow anymore due to the natural decline of the pebble flux.
On the contrary, for $\alpha = 10^{-5}$ the pebble accretion rate is extremely high, and $\Miso$ is reached within the first 46'000 yrs of disc evolution. 
This means that most of the pebbles that continue drifting through the planet's orbit cannot contribute to the planets' growth. 
 
\begin{table*}
\caption{Pebble flux and isolation mass for the nominal set-up ($Z_0 = 0.01$, $a=0.2$ au)} 
\label{TabPflux}      % is used to refer this table in the text 
\begin{center}                                    % used for centering table
\begin{tabular}{l c c c c c c }           % centered columns (4 columns)
\hline                                   % inserts single horizontal line

 $\alpha$ &  $\Md$ [M$_{\odot}$]  & Final $\Mcore$ [\ME]& Time when $\Miso$ [Myr] & Time when pebble &  Integrated  &  $\epsilon$ [\%] \footnotemark  \\ %
  &  & & is achieved &  flux is negligible [Myr] \footnotemark & pebble flux [\ME] & \\
 
 \hline                      % inserts double horizontal lines
 		 &0.01&  0.037 & never  & 0.44 & 0.20 & 13.5 \\
$10^{-3}$	&0.03 & 0.026 & never & 1.13 & 0.56 & 2.86 \\
		& 0.06& 0.017 & never &1.92 & 1.00 & 0.7   \\
		& 0.1& 0.015 & never & 2.80 & 1.66  & 0.3 \\
\hline
 	 	& 0.01 & 0.25 & never & 0.88 & 7.77 & 3.09 \\
$10^{-4}$	 &0.03 & 0.87 & never & 1.16& 23.4 & 3.68 \\
		& 0.06& 1.38 & 3.70 & 2.23 & 45.7 &  3.00 \\
		& 0.1&  1.95 & 3.73 & 3.71 & 74.2 & 2.61 \\
\hline
		 & 0.01 & 1.76  & 0.069 & 2.43 & 25.9 & 6.76 \\
 $10^{-5}$	  &0.03 & 2.58 & 0.047 & 2.31 & 76.2 & 3.37 \\
		& 0.06& 4.60 & 0.046 & 2.32 & 157 & 2.92 \\
		& 0.1&  7.30 & 0.047 & 2.17 & 267 & 2.73 \\

\hline                                             %inserts single line
\end{tabular}
\end{center}
\footnotemark[1]{$\epsilon$ is the efficiency of pebble accretion, defined as the amount of pebbles accreted by the planet over the integrated pebble flux at the position of the planet \citep{Guillot14b, LJ14}}\\
\footnotemark[2]{Negligible means that the pebble flux falls below $10^{-8}$ \ME/yr.}\\

\end{table*}

%--------------------------- TABLA Pebble flux -----------------------------------------------------------

Let us analyse now the results of the nominal set-up (figure \ref{fig_nominal_res}) for each choice of $\alpha-$viscosity.
As we already explained, for the cases where $\alpha = 10^{-3}$, the turbulence fragments the pebbles into small dust particles ($\sim10-100 \, \mu$m), which translates into almost negligible pebble accretion rates.
This is why these cores remain basically as lunar-mass embryos. However, it is interesting to note that contrary to the other viscosities, the core mass increases slightly more for lower disc masses. 
In fact, that little growth reported in the table happens just after photoevaporation opens a gap in the mid-disc and the gas of the inner disc starts to disappear. 
At that point the decrease in the gas surface density allows the small particles that are within the Hill sphere of the embryo (there are still particles around since due to their small size they hardly drift) to increase their stokes number, fostering pebble accretion for one blink of time. For lower disc masses this happens earlier,  when there are typically more pebbles around. This explains the trend in the final embryo masses. 
This effect practically does not occur for other viscosities because by the time the Stokes number starts to increase due to the gas depletion, the pebble isolation mass is already attained.

Interestingly, the stall of growth for $\alpha = 10^{-3}$ occurs despite of the initial dust-to-gas ratio, and also for migrating planets, as we show in the following sections.
We can therefore conclude that discs with intermediate and high turbulence  ($\alpha \geq 10^{-3}$) cannot form planets by pebble accretion inside the water ice line. 

For discs where $\alpha = 10^{-4}$, the planet mass varies from 0.25 \ME \, for the lowest initial disc mass ($\Md$ =0.01 $\Msun$) to a 1.95 \ME \,for the initial highest one ($\Md$ =0.1 $\Msun$). 
For the two lowest disc masses, the pebble flux stops before the pebble isolation mass is reached because the disc runs out of pebbles. 
For the two highest disc masses, the pebble isolation mass is reached close to the disc dispersal, when $\Miso$ is low. 
This shows the importance of considering the physical dimensions of discs, because the final core mass of planets formed by pebbles can be determined by the halt in the pebble supply and not solely by the attainment of the pebble isolation mass.

For discs where  $\alpha = 10^{-5}$, the flux of pebbles and the pebble accretion rates are very high, so the pebble isolation mass is rapidly achieved. 
This sets the final core mass shown in Fig. \ref{fig_nominal_res} for this $\alpha$.
These cases of extremely low viscosity show the formation of two types of planets: either rocky super-Earths of masses below 5 \ME, or ``gas-dwarfs'' for the two largest initial disc masses. 
We use the term gas-dwarf to refer to planets whose envelopes contribute greatly to the total mass, i.e, $\fhhe \sim 50\%$. 
These ``gas-dwarfs'' are very interesting objects, they have a gas-to-core ratio of almost unity. 
The reason why these planets do not explode into gas giants is the opening of a gap at low masses caused by the low disc viscosity (see Eq.\ref{eqCrida}) and the low disc aspect ratio at short orbital distances. This halts gas accretion at $\Mp \gtrsim$ 8 \ME\, (see dashed-purple line of Fig.\ref{fig_pebAcc}, bottom).   
Since these planets occur only at high disc masses and extremely low (probably unrealistically low) disc viscosities, they should rarely occur in nature.

We close this section by analysing the efficiency of pebble accretion for the nominal set-up. 
Such efficiency can be defined as the ratio between the final core mass and the integrated pebble flux at the end of the simulation \citep{Guillot14b, LJ14}.
This is shown in the last column of Table \ref{TabPflux} and its represented by $\epsilon$. We note that the integrated pebble flux is always smaller than the initial amount of dust in the disc. Indeed, many of the outer pebbles do not reach the planet's orbit, for instance, once photoevaporation opens a gap in the mid-disc, the inner disc becomes detached from the outer one and the pebbles coming from far cannot reach anymore a planet growing in the inner disc. In addition, the different $\alpha$ set different mean sizes for the dust/pebbles, which translates into very different drift speeds. The larger the $\alpha$, the smaller the grains and the less they drift. This explains the trend of the reduction of the integrated pebble flux with increasing $\alpha$. 
The low pebble accretion efficiencies that we find are in general agreement with the results of \citet{Ormel18}.
For the lowest disc masses of the cases with $\alpha  =10^{-3} $ our efficiencies are larger, but, as explained above, this is originated by photoevaporation, which is not taken into account in \citet{Ormel18}.

\subsection{Dependence on the initial dust-to-gas ratio or stellar metallicity}
Exoplanets with periods shorter than 100 days exist around stars with a wide range of stellar metallicities, spanning in [Fe/H] from -0.5 to +0.5 dex \citep[e.g.][]{Buchhave15}.   
This motivates us to study rocky planet formation for different disc metallicities, which presumably are representative of the star metallicity at the time of the system's formation.

\begin{figure*}
\begin{center}
	\includegraphics[width=0.9\textwidth]{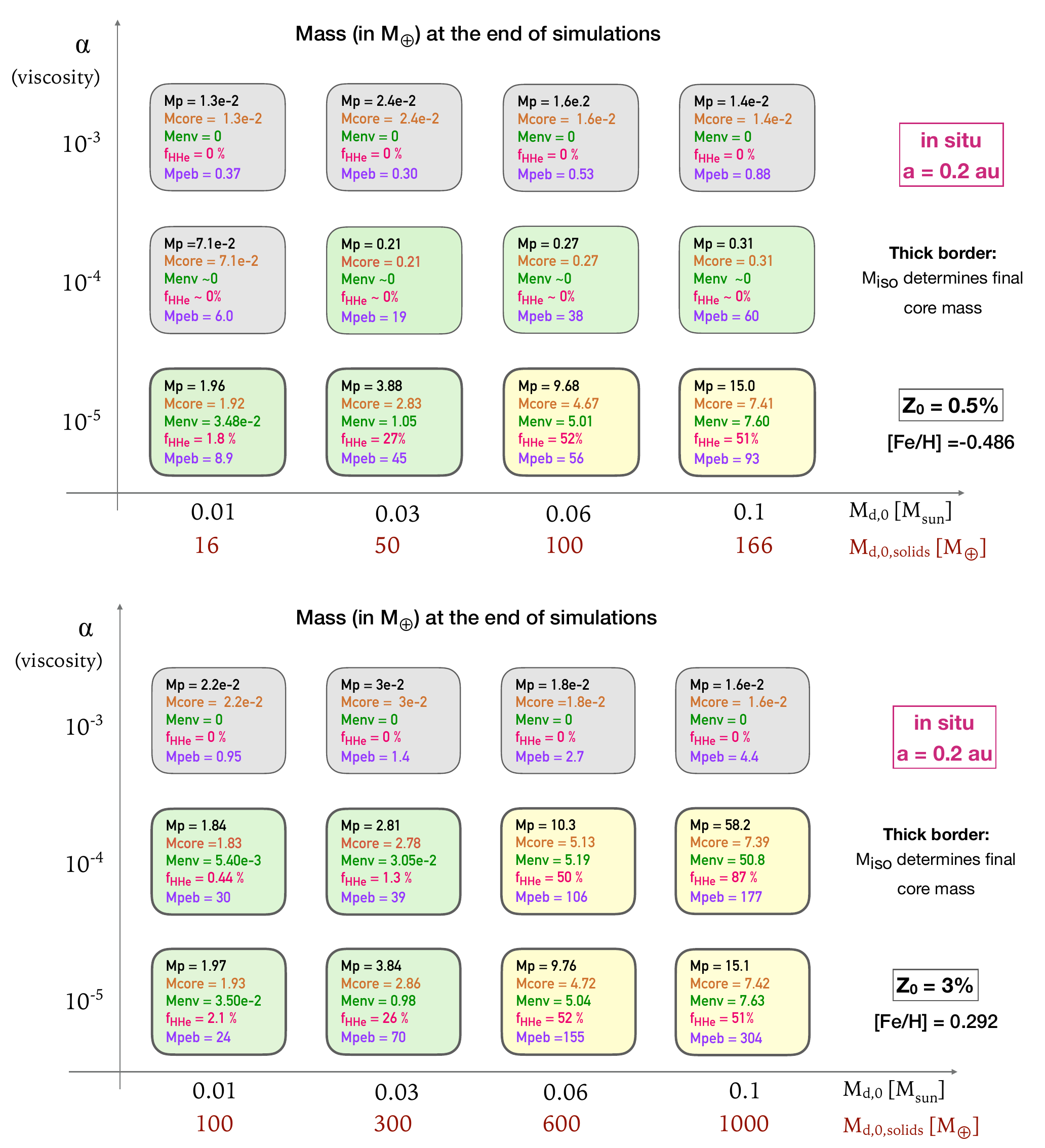}
	\caption{Same as Fig.\ref{fig_nominal_res} but for a low ($Z_0$  = 0.005) and high  ($Z_0$  = 0.03) initial dust-to-gas ratio.}
\label{fig_Z0}
\end{center}
\end{figure*}

We repeated our runs for an initial dust-to-gas ratio of $Z_0 = 0.005$, which taking the protosolar abundances of $Z_{\odot} = 0.0153$ \citep{Lodders09}, would correspond to a metallicity of [Fe/H] $\approx -0.486$. 
This is an extremely low stellar metallicity, representative of the lowest values of stars-hosting planets \citep{Ghezzi2010}.
The output of this setup is summarised in Fig.\ref{fig_Z0}.   
We see that in this case, $\alpha = 10^{-3}$ continues leading to practically no growth. For $\alpha = 10^{-4}$, three out of four cases lead to the Mars-mass embryos.
The case of extreme low viscosity of $\alpha = 10^{-5}$ is very similar to the nominal one. This is a trend for any disc metallicity. 
The reason behind it is the quick attainment of the pebble isolation mass, which does not vary by changing $Z_0$ because it only depends on the disc's aspect ratio. 

We also repeated the simulations for a high initial dust-to-gas ratio of $Z_0 = 0.03$, corresponding to [Fe/H] $\approx 0.292$. 
Actually our results do not change for metallicities higher than this.
The results are summarised in the lower panel  of Fig.\ref{fig_Z0}. 
Again, when $\alpha = 10^{-3}$ planetary seeds do not grow beyond moon-mass. 
For  $\alpha = 10^{-4}$, the two lowest disc masses form terrestrial planets. 
For the two highest disc masses, the cores grow larger than 5 \ME \, and accrete substantial envelopes.

\subsection{Terrestrial planet formation for different embryo locations}
We performed the same simulations assuming different initial locations for the embryo. 
We first analyse two other cases of in-situ growth, and then one case where the embryo is allowed to migrate from just inside the water ice line.

\subsubsection{In-situ growth}\label{sec_a01_1}
\begin{figure*}
\begin{center}
	\includegraphics[width=0.87\textwidth]{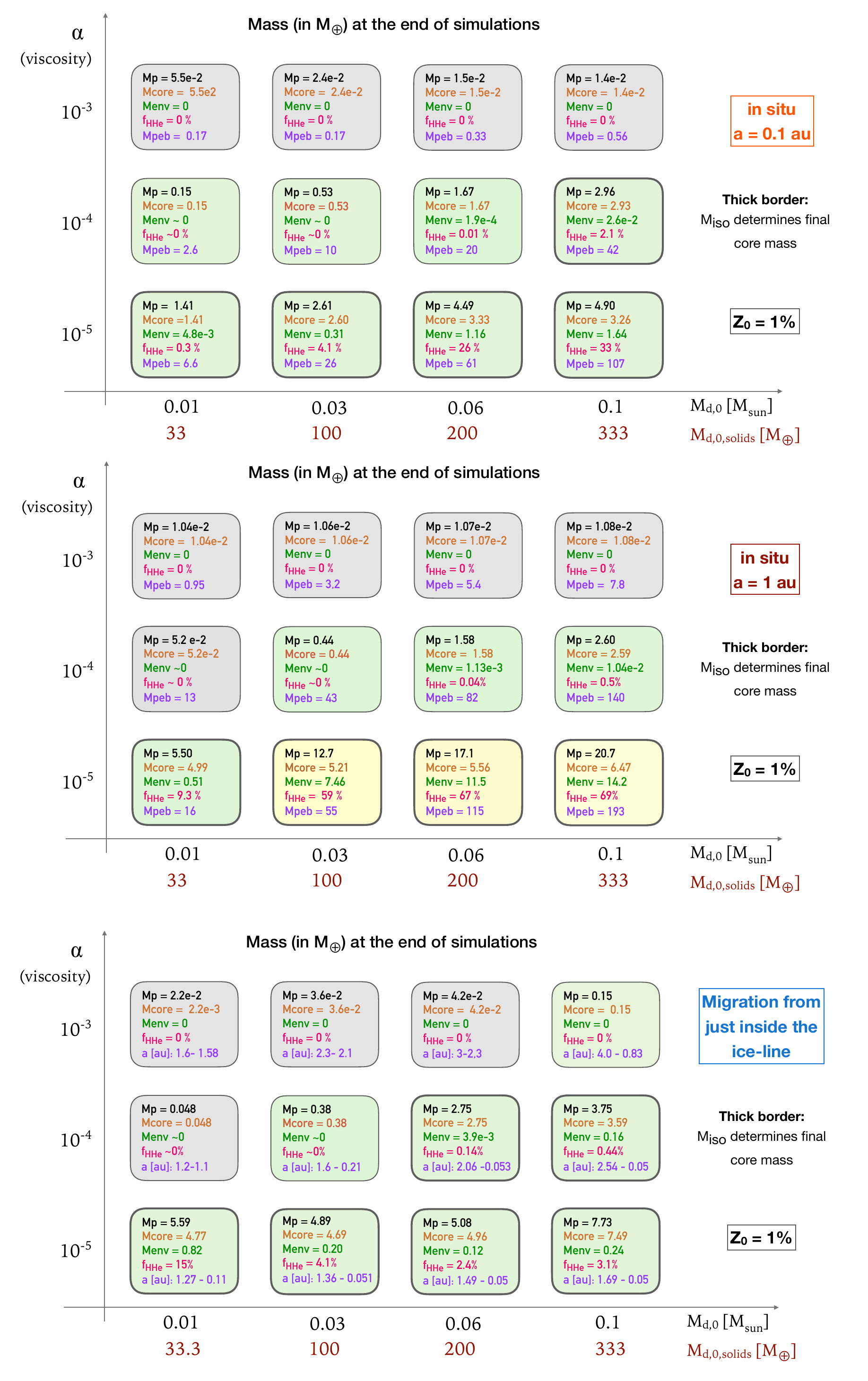}
	\caption{Same as Fig.\ref{fig_nominal_res} but for the planet growing at different locations. The top and middle panels assume in-situ growth at 0.1 au and 1 au, respectively. 
	For the bottom panel planets start at 0.2 au from the ice line (inside it) and are allowed to migrate. The range of semi-major axis of each box (in purple) indicates the initial and final planet location.}
\label{tab_semiejes}
\end{center}
\end{figure*}

When repeating the calculations shown in figure \ref{fig_nominal_res} for $a=0.1$ au and $a = 1$ au, we find that the typical outcome of terrestrial planets for $\alpha = 10^{-4}$ also holds for different semi-major axis as shows Fig.\ref{tab_semiejes}.
The reason for the results not to be very sensitive to the embryo location (as long as it lays within the ice line), is that the pebble surface density tends to increase for shorter orbital periods (e.g. bottom-left panel of Fig.\ref{fig_GasDisc}), while the opposite is true for the Stokes number (Fig.\ref{fig_pebblesize}). 

\begin{figure}
\begin{center}
	\includegraphics[width=\columnwidth]{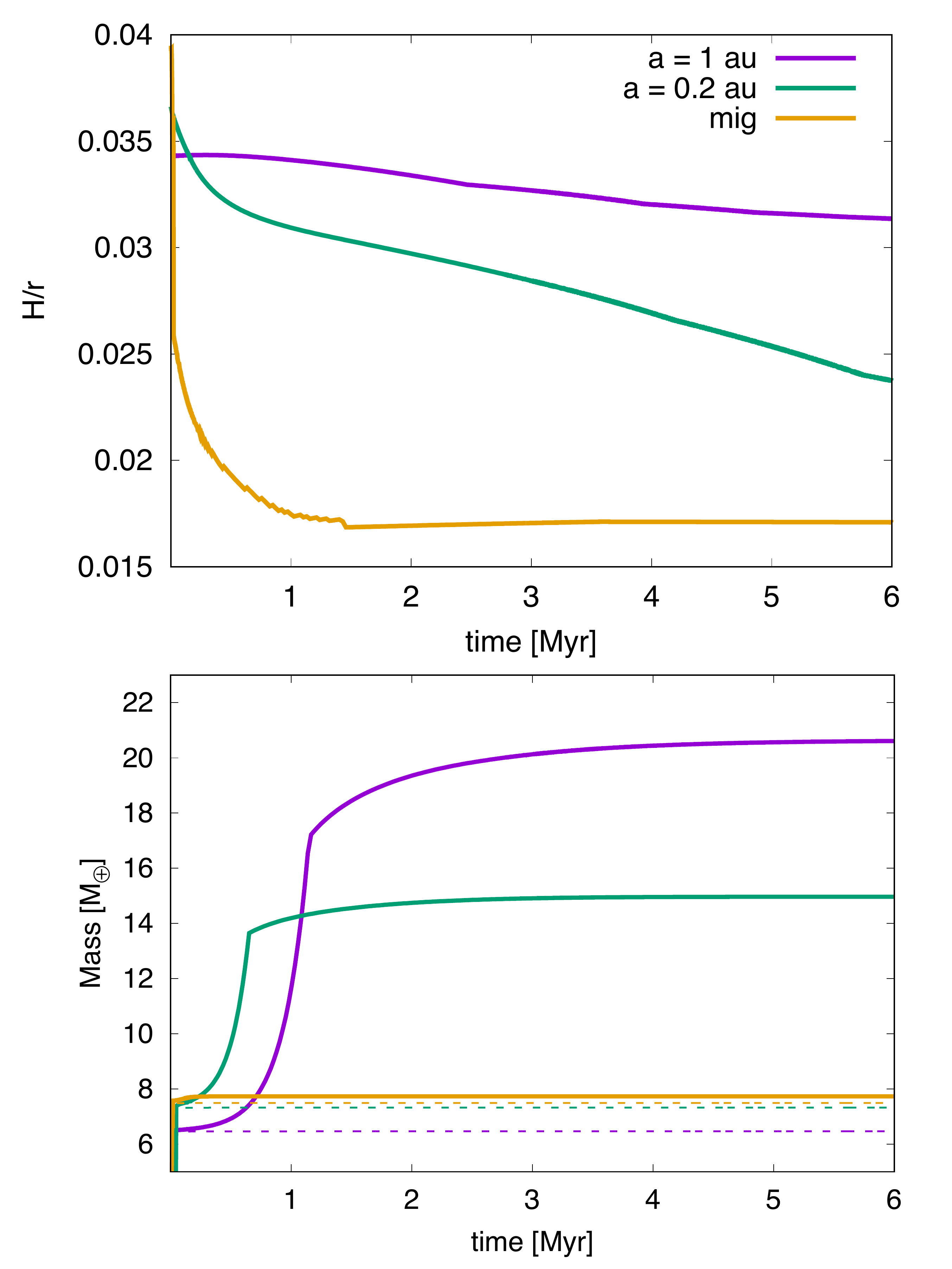}
	\caption{Evolution of aspect ratio at the planet's location (top), and planet mass (bottom) for three cases where $\Md = 0.1 \, \Msun$ and $\alpha = 10^{-5}$. The dashed lines indicate the evolution of $\Mcore$ and the solid lines the evolution of $\Mp$. The planet's location is specified in the legend. For the migrating case, the evolution of $a$ is shown in the next figure.}
\label{fig_aspect_ratio}
\end{center}
\end{figure}

For $\alpha = 10^{-5}$, we note that the mass fraction of H-He increases for larger semimajor-axis. 
This is related to the increase of the aspect ratio for longer orbital periods, as shows the bottom-left panel of Fig.\ref{fig_GasDisc}. The larger the $H/r$, the more massive a planet can grow without opening a gap.
We illustrate the link between the evolution of the aspect ratio and planet mass for a planet growing at $a=1$ au, $a=0.2$ au and a migrating case in Fig.\ref{fig_aspect_ratio}. All these cases reach $\Mcore \sim 7$ \ME, but despite of having the same viscosity of $\alpha = 10^{-5}$ , the final mass reach different values due to the difference in aspect ratio.

 \subsubsection{Migration from the ice line}\label{sec_mig}
\begin{figure}
\begin{center}
	\includegraphics[width=\columnwidth]{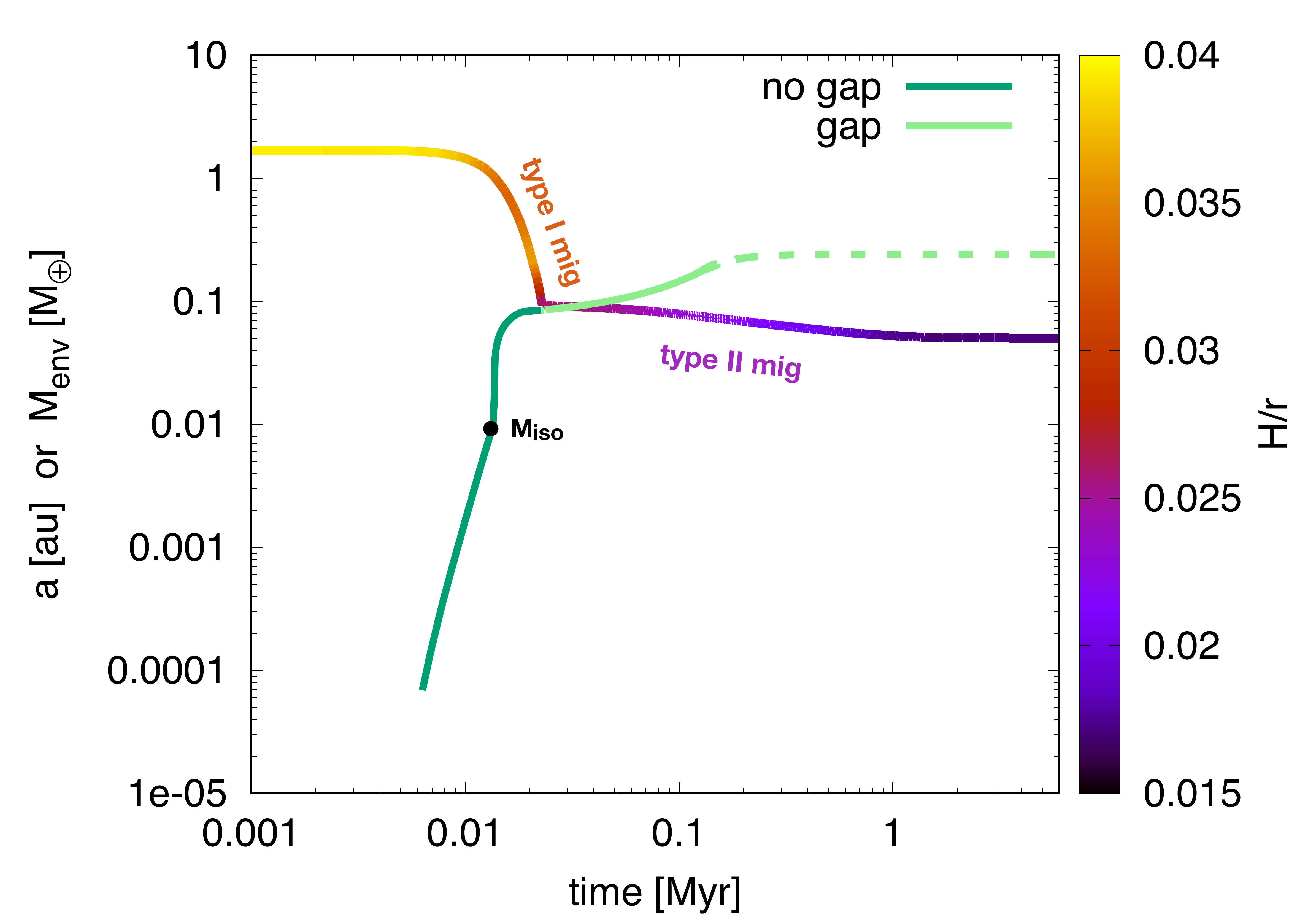}
	\caption{Evolution of semimajor axis (yellow-purple) and envelope mass (green curve) for the migrating case of Fig.\ref{fig_aspect_ratio}. The time when the isolation mass is reached is marked with a black dot. The planet transitions from type I to type II migration when the gap opens (light green curve). The dashed curves indicate that gas accretion follows \citet{Tanigawa07}.}
\label{fig_migration}
\end{center}
\end{figure}

Finally, we also tested the possibility of planetary migration. For these runs the embryos start at 0.2 au from the ice line (inside the ice line to keep them dry by construction). The initial ice line location depends on the mass of the disc and on the value of the $\alpha$ parameter, so it is different for each disc. All the planets that migrate and reach a planet mass larger than 1 \ME, migrate fast and park close to the inner edge of the disc, where a pressure bump exists due to the zero torque boundary condition at the inner edge of the gas disc.
We find the same trends as before, with the exception that more rocky-dominated planets result for $\alpha = 10^{-5}$. 
Indeed, for the highest disc mass, a rocky planet with $\Mcore = 7 \, \Mearth$ is formed. 
The halt of gas accretion for this massive core is related to the low disc's scale height at shorter orbital periods, as we explained above.
In Fig.\ref{fig_migration} we show that soon after attaining isolation mass, this planet opens a gap at $\Menv \approx 0.1 \,\Mearth$. At this time the planet switches from type I to type II migration, moving slower until parking at the inner edge of the disc. We note that gas accretion stops earlier than this, due to the low accretion  rates with gap opening given by \citet{Tanigawa07} (dashed-green lines).
We study the migration scenario more in-depth in an accompanying letter \citep[][hereafter, Paper II]{Venturini2020Letter}Paper II.

\subsection{Planetary radii}
\begin{figure}
\begin{center}
	\includegraphics[angle=270, width=\columnwidth]{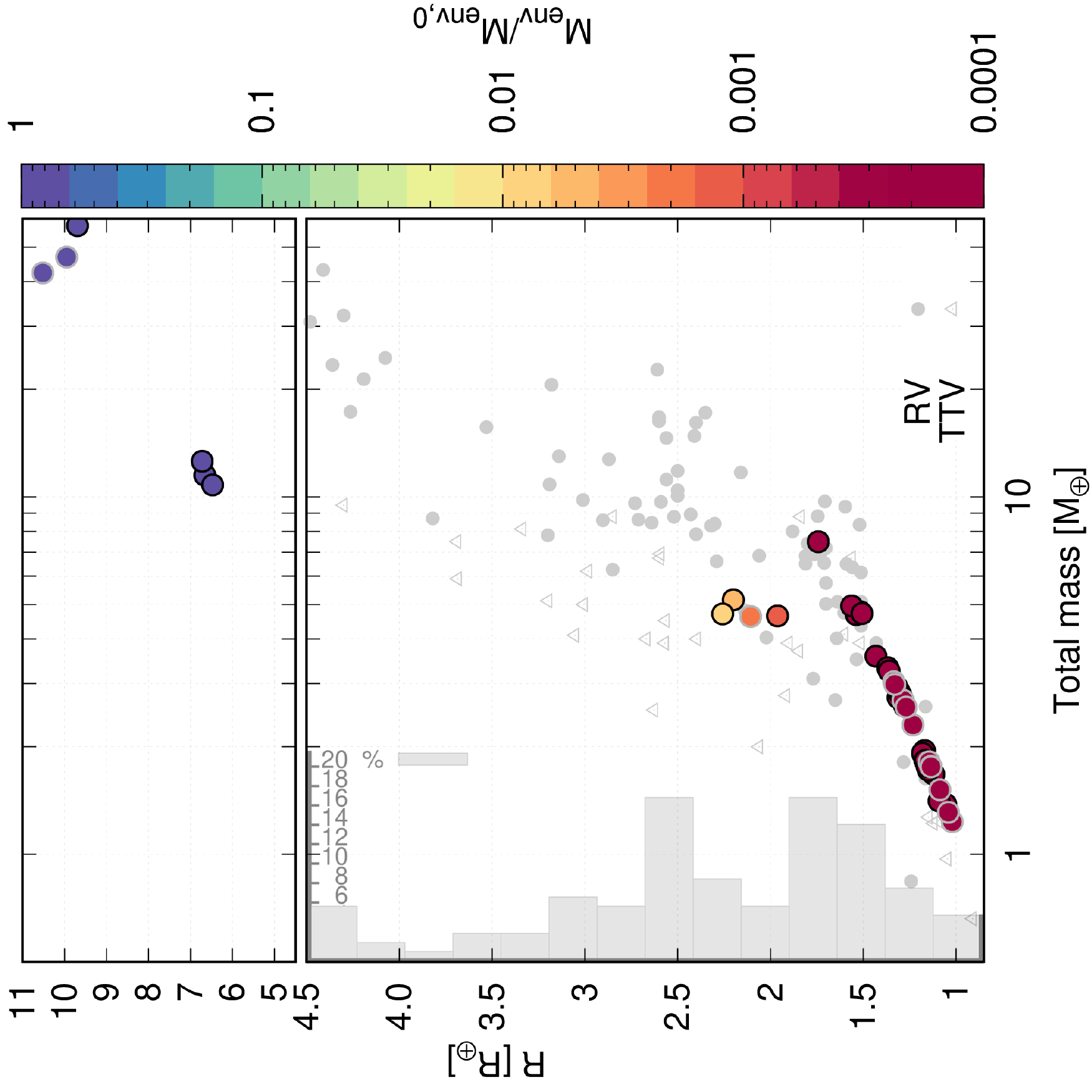}
	\caption{Final mass-radius for all the planets with P < 100 days and $\Mcore \geq 1 \, \Mearth$, represented by black-border colored circles. The grey-border colored circles represent the planets computed with low dust opacity (Sect.\ref{sec_lowOpac}). The colors show the final envelope mass relative to the initial one. The final mass of the envelope and radii are computed following Sec. \ref{sec:Christoph}. The grey small dots/triangles represent the RV/TTV planets used by \citet{Zeng19} (from their Fig.2 and Table S1) and were used to construct the grey histogram on the left, which depicts the percentage of planets as a function of their radius.} 
\label{fig_MR}
\end{center}
\end{figure}

We are interested in understanding how silicate-pebble accretion contributes to the formation of super-Earths/mini-Neptunes. 
Out of the 60 planets we form with orbital period below 100 days (i.e, considering all results except in-situ formation at $a=1$ au); 30 accrete a core more massive than 1 $\Mearth$ and, for those, we show the final mass-radius plot, constructed by following Sec. \ref{sec:Christoph} in Fig.\ref{fig_MR}. Of those 30 planets, 23 finish with radii below 1.8 $\RE$, that is, most of the super-Earths we form contribute to the first peak of the Kepler size distribution. 
One planet of those 30 falls exactly in the valley, with radius of 1.96 $\RE$, and 2 fall in the second peak with $\Rp \approx 2.2-2.3$ $\RE$. These last three cases correspond to planets with $\Mcore \approx 4.6-5.1 \, \Mearth$ which had originally an envelope of equal mass and lost most of it to end up with a final envelope mass of 5.5$\times10^{-3}$ \, \ME\, for the planet in the valley and 2-3 $\times10^{-2}$\, \ME\, for the two planets in the second peak. 

Interestingly, 3 of these 30 planets have radius between 6.5 and 6.7 $\RE$. These objects have cores of approximately 7 \ME, started with envelopes of equal mass and finished with envelopes ranging 3-5 \ME. These 3 cases are planets originating in extremely low viscosity discs ($\alpha = 10^{-5}$), and hence, are expected to be rare.
Finally, the only true gas giant we formed ($\Mcore = 7.4 \,\Mearth$,  $\Menv  = 51 \,\Mearth$, $\alpha = 10^{-4}$) loses barely $\sim$1\% of its original envelope and finishes with a radius of 9.7 $\RE$.

\begin{table*}
\caption{All planets with $\Mcore\geq 1 \, \Mearth$ which retain some H-He after 5 Gyr of evaporation driven mass-loss.} 
\label{TabMR}      % is used to refer this table in the text
\centering                                      % used for centering table
\begin{tabular}{c c c c c c c c c }           % centered columns (4 columns)
\hline                                   % inserts single horizontal line
 Z$_{0}$ & $\alpha$ &  $\Md$ [M$_{\odot}$]  & $a$ [au] & $\Mcore$ [\ME] & $M_{\rm env, ini}$ [\ME] & $M_{\rm env, final}$ [\ME] & \% $\Menv \,\text{lost}$  &  $\Rp \,[\RE]$ \\
 \hline                      % inserts double horizontal lines
0.005 & $10^{-5}$ & 0.06 & 0.2 & 4.67 & 5.02 & $3.81\times10^{-2}$ & 99.24 & 2.26 \\
0.005 & $10^{-5}$ & 0.1 & 0.2 & 7.41 & 7.60 & 5.18 & 31.91 & 6.72 \\
0.01 & $10^{-5}$ & 0.06 & 0.2 & 4.64 & 4.99 & $5.47\times10^{-3}$ & 99.89 &  1.96 \\
0.01 & $10^{-5}$  & 0.1  & 0.2 & 7.32 & 7.64 & 4.18 & 45.32 & 6.66 \\ 
0.03 & $10^{-4}$  & 0.06 & 0.2 & 5.13 & 5.19 & $2.25\times10^{-2}$ & 99.56 & 2.20 \\
0.03 & $10^{-5}$  & 0.1  & 0.2 & 7.42 & 7.63 & 3.39 & 55.51 & 6.48 \\
0.03 & $10^{-4}$ & 0.1 & 0.2 & 7.39 & 50.79 & 50.07 & 1.42 & 9.69 \\
\hline
0.01 & $10^{-4}$ & 0.1 & 1.0 & 2.58 & 0.01 & $2.52\times10^{-3}$ & 75.68 & 1.51 \\
0.01 & $10^{-5}$ & 0.01 & 1.0 & 4.98 & 0.51 & 0.47 & 7.20 & 3.17 \\
0.01 & $10^{-5}$ & 0.03 & 1.0 & 5.21 & 7.46 & 7.36 & 1.39 & 7.11 \\
0.01 & $10^{-5}$ & 0.06 & 1.0 & 5.56 & 11.54 & 11.47 & 0.54 & 7.63 \\
0.01 & $10^{-5}$ & 0.1 & 1.0 &  6.47 & 14.17 & 14.11 & 0.37 & 7.68 \\
\hline                                             %inserts single line
\end{tabular}
\end{table*}

\begin{table*}
\caption{Low dust opacity cases 
($f_{\rm dust}= 0.01$). The columns are as in Table \ref{TabMR}, but considering all the cases run with a low dust opacity.$^4$} %\footnotemark}}
\label{TabMR-low-opacity}      % is used to refer this table in the text
\begin{center}                                    % used for centering table
\begin{tabular}{c c c c c c c c c }           % centered columns (4 columns)
\hline                                   % inserts single horizontal line
 Z$_{0}$ & $\alpha$ &  $\Md$ [M$_{\odot}$]  & $a$ [au] & $\Mcore$ [\ME] & $M_{\rm env, ini}$ [\ME] & $M_{\rm env, final}$ [\ME] & \% $\Menv \,\text{lost}$  &  $\Rp \,[\RE]$ \\
 \hline                      % inserts double horizontal lines
0.01 & $10^{-5}$ & 0.01 & 0.2 & 1.76 &  4.70 & 0.00 & 100 & 1.14 \\
0.01 & $10^{-5}$ & 0.03 & 0.2 & 2.58 &  5.96 & 0.00 & 100 & 1.27 \\
0.01 & $10^{-5}$ & 0.06 & 0.2 & 4.61 &  7.28 & 0.02 & 99.77 & 2.11 \\
0.03 & $10^{-4}$ & 0.1  & 0.2 & 6.79 & 64.94 & 64.49 & 0.68 & 9.92 \\
0.03 & $10^{-4}$ & 0.06 & 0.2 & 4.85 & 42.56 & 42.08 & 1.12 & 9.95 \\
0.03 & $10^{-4}$ & 0.03 & 0.2 & 2.69 & 19.16 & 0.00 & 100 & 1.29 \\
0.03 & $10^{-4}$ & 0.01 & 0.2 & 1.81 &  1.85 & 0.00 & 100 & 1.15 \\
0.01 & $10^{-4}$ & 0.06 & 0.2 & 1.23 &  1.30 & 0.00 & 100 & 1.02 \\
0.01 & $10^{-4}$ & 0.1  & 0.2 & 1.76 &  2.51 & 0.00 & 100 & 1.13 \\
0.01 & $10^{-5}$ & 0.01 & 0.1 & 1.31 &  3.26 & 0.00 & 100 & 1.04 \\
0.01 & $10^{-5}$ & 0.03 & 0.1 & 2.30 &  3.99 & 0.00 & 100 & 1.23 \\
0.01 & $10^{-5}$ & 0.06 & 0.1 & 2.98 &  5.66 & 0.00 & 100 & 1.33 \\ 
0.01 & $10^{-5}$ & 0.1  & 0.1 & 3.04 &  7.69 & 0.00 & 100 & 1.33 \\
0.01 & $10^{-4}$ & 0.06 & 0.1 & 1.52 &  1.50 & 0.00 & 100 & 1.08 \\
0.01 & $10^{-4}$ & 0.1  & 0.1 & 2.77 & 41.81 & 39.58 & 5.33 & 10.52 \\
\hline                                             %inserts single line
\end{tabular}
\end{center}
\footnotemark[4]{Only two of the cases appear in Table \ref{TabMR} for the nominal opacities, and those are shown in the same row location (rows 3 and 5). The remaining cases should be checked against Figs. \ref{fig_nominal_res}, \ref{fig_Z0}, and \ref{tab_semiejes} for comparison to the nominal opacities.}
\end{table*}

\section{Discussion} \label{sec_discussion}
\subsection{Implications for the Kepler size distribution}
By studying planet formation by dry pebble accretion with pebbles growing self-consistently from initial micro-meter dust, we found that out of 72 simulations, 35 cases lead to the formation of rocky-dominated planets (green squares in Figs. \ref{fig_nominal_res}, \ref{fig_Z0} and \ref{tab_semiejes}), of which 34 remain as naked cores after 5 Gyr of evolution. The masses of these bare cores range typically from Mars to 4 \ME. These planets contribute to the first peak of the Kepler size distribution. The only case of these rock-dominated planets that retains some atmosphere after evolution is a core that reaches 4.98 \ME\, at $a= 1$ au (see Table\ref{TabMR}), which, due to the low irradiance, retains 93\% of the original 0.51 \ME\, envelope, rendering a planet radius of 3.17 \ME. 
Especial cases with $\Mcore >4 \, \Mearth$ are migrating planets in extremely low viscosity discs, which can remain as bare cores even for $\Mcore$ = 7.5 \ME, as long as the planet ends with an orbital period as short as 4 days.

Out of the total 72 cases, only three planets within 100-day orbital period fall in the gap/second peak of the size distribution (see Table \ref{TabMR}). Interestingly, these planets started as gas-dominated objects ( or ``gas-dwarfs", i.e, having more than 50\% in mass of H-He), but lost more than 99\% of their envelope due to evaporation. Equally interesting, cores with $\sim$7 \ME \, instead of 5 \ME \, lose considerably less H-He ($\sim$50\%), yielding a planet radius of 6-7 $\RE$ (see Fig.\ref{fig_MR} Table \ref{TabMR}). Several of this so-called
`sub-Saturns' have been found \citep{Petigura18}, and their origin poses challenges \citep{Millholand20}.

Finally, we note that our silicate-pebble accretion model does not produce any planet with size between 2.5 and 6 $\RE$. Planets with sizes between 2.5 and 4 $\RE$ are known to be very common \citep[e.g][also clear from the observed exoplanets in Fig.\ref{fig_MR}]{Petigura18}. Our results suggest that those planets might not be dry. We address the origin of those objects in Paper II.

\subsection{Dependence on the dust opacity}\label{sec_lowOpac}
We closed last section remarking that with our pure silicate-pebble accretion model we cannot form planets ending up with radii between 2.5 and 6 $\RE$. This could depend on our choice of opacities. In Sect.\ref{sec_GasAtt} we justified our choice of ISM opacities on the grounds of maximising the possibility of forming bare rocky cores. However, grain growth within the envelope could reduce the dust opacity by a factor of $\sim$100 compared to ISM values \citep{Mordasini14, Movshovitz08}. Thus, reduced opacities can lead to larger envelopes for equal core masses and hence could contribute to populate this range of $\Rp \sim 2.5- 6 \, \RE$. 

To evaluate this possibility, we repeated 15 of our original simulations for those cases that would certainly lead to $\Rp<6.5 \, \RE$. The outcome of this test is presented in Table \ref{TabMR-low-opacity} and is shown as well as the grey-border circles of Fig.\ref{fig_MR}.

Let us first note that a planet that ended up with $\Mcore \approx 3  \, \Mearth$ has now an envelope mass after formation of $\Menv \approx 20  \, \Mearth$ (sixth line of Table \ref{TabMR-low-opacity}), contrary to the nominal case where $\Menv \approx 0.03  \, \Mearth$ (Fig.\ref{fig_Z0}). This is due to the rise of the gas accretion rate promoted by the lower opacities \citep{Ikoma00}. Nevertheless, despite of the thick envelope that the planet acquires with the low dust opacity, the envelope is completely removed by photoevaporation due to the low core gravity and relatively low overall planet gravity. 
The fact that similar cores that accreted very different envelopes can yield to the same planetary radius after evaporation, was pointed out by \citet{Mordasini20} as an obstacle to constrain the dust opacities from radii measurements.  
For larger core and envelope masses, the massive envelope accreted as a consequence of the low opacity can be retained after photoevaporation, making now a planet with $\Mcore \approx 5  \, \Mearth$ to have $\Rp \approx 10 \, \RE$ (fifth row of Table \ref{TabMR-low-opacity}), compared with $\Rp = 2.2 \, \RE$ of the nominal case (fifth row of Table \ref{TabMR}).
A similar result was found by \citet{Ogi20}, who assumed the envelopes as dust-free. The planets that they form retain too massive envelopes, reason for which they find too inflated planetary radii compared to the Kepler size distribution. 

Overall, we note that the lower dust opacities also do not produce any planet in the size range of 2.5-6 $\RE$ (Fig.\ref{fig_MR} and Table \ref{TabMR-low-opacity}). Either the thick envelopes are blown away by photoevaporation, leaving in the best case one planet in the second peak; or are retained for sufficiently massive planets ($\Mp \gtrsim 30 \,\Mearth$), yielding planetary radii above 6 $\RE$.
In Paper II, where we perform a larger number of simulations, we find some planets with $\Rp \sim 2.5-6 \,\RE$, but they are typically migrated planets from beyond the ice line and therefore, ice-rich.

\subsection{Dependence on the disc aspect ratio}
One of our main findings is that pure rocky planets form with low mass, typically below 5 $\Mearth$. 
The main property that sets this maximum mass is the pebble isolation mass \citep{Lambrechts14, Ormel17, Liu19}, which depends strongly on the disc's aspect ratio (Eq.\ref{MisoEq}), and therefore, on the disc model.
The low values of the pebble isolation mass that we find in the inner part of the disc correspond to aspect ratios of $\sim$0.02-0.03 within 1 au (Fig.\ref{fig_GasDisc}).
The disc's scale height depends on the thermodynamical state of the disc, which is determined by the heating and cooling processes at operation. We consider that the disc is heated by irradiation from the central star, on top of the heat produced by viscous accretion. The grain opacity is responsible for disc's cooling, and it is taken as the one of \citet{BL94}, which corresponds to micro-meter size grains. This is an upper bound for the grain opacity \citep{Savvidou2020} \footnote{Although porosity could increase the dust opacity by a factor of $\sim$2 for $800 \lesssim T \lesssim 1200$ K \citep{Semenov03}.} and therefore maximises the disc scale height and the pebble isolation mass.

Despite of the dependence of $\Miso$ on the disc model, other studies find similar values of pebble isolation mass than us. For instance, \citet{Liu19} use a simple steady state disc solution, and claim that for an embryo's initial location within 1 au, the planets "grow moderatly up to a few Earth masses". \citet{Bitsch19a}, also via a steady state disc solution, find that "the disc structure leads normally to very small aspect ratios
(around 2-2.5\%) in the inner disc (within 1 au) at late times, which results in very small core masses of 1.6-3.0 Earth masses". Passive discs \citep{Bitsch19b} or discs with lower grain opacity \citep{Savvidou2020} would yield to even lower values of disc scale height, which reinforces our conclusion of pure rocky planets not being more massive than $\approx$ 5 $\Mearth$ when formed by pebble accretion (neglecting post-disc N-body interactions).

\subsection{Formation of rocky planets by planetesimal accretion}
In this work we considered rocky planet formation driven solely by pebbles. Planetesimals might also form and contribute to the planetary growth \citep[see][and references there in]{Venturini2020}, although in most of our studied cases ($\alpha \geq 10^{-4}$), the condition of streaming instability to form planetesimals \citep[e.g][]{Drazkowska16} is not fulfilled inside the water ice line. 
Nevertheless, other mechanisms that produce planetesimals could operate. For instance, \citet{Lenz19} propose that hydrodynamical or magnetic instablities could create turbulent structures along the entire disc that could clump solids triggering planetesimal formation. They proposed a parameterised model in which planetesimals form by a local mechanism that converts a certain fraction of the pebble flux into planetesimals. Basically, when the pebble flux becomes greater than a critical mass value over the lifetime of the turbulent structure, some fraction of pebbles are transformed into planetesimals along the length-scale of the turbulent structure. Thus, this mechanism is directly related to the local pebble flux instead of  to the local dust-to-gas ratio as the streaming instability mechanism. \citet{Lenz19} showed that a high $\alpha$-viscosity parameter ($\alpha= 10^{-2}$) does not allow the formation of planetesimal in the inner part of the disc, and that a moderate value ($\alpha= 10^{-3}$) generates steep planetesimal surface density profiles. Recently, \citet{Voelkel20} introduced a model of dust growth and evolution, and the planetesimal formation model of \citet{Lenz19} in a  planet population synthesis model \citep{Mordasini2018}. They found that the steep planetesimal surface density profile generated by the pebble flux mechanism of \citet{Lenz19} increases the rate of formation of terrestrial planets, super-Earths
and gas giants by the accretion of 100 km planetesimals in the inner region of the disc.

\subsection{Growth by collisions after the disc's dispersal}
\citet{Ogi20} and \citet{Lambrechts19} simulated the formation of super-Earths by both pebble accretion and N-body interactions.
These studies show that the main driver of growth during the disc phase is the accretion of pebbles. 
Indeed, since pebble accretion is inefficient, meaning that most of the pebbles are not used to grow a planetary core (as we show in Table \ref{TabPflux}); 
when many embryos grow together there are enough pebbles for all and they all grow at similar rates 
\citep{Ogi20, Lambrechts19}. After disc dispersal, a super-Earth typically undergoes one or two giant collisions \citep{Ogi20}. 
Hence, if producing pure rocky planets of $\Mp \gtrsim$ 5 \ME \, is atypical, pure rocky planets exceeding 10 \ME \, should also be unusual. This seems to be the case when one analyses the exoplanets discovered so far and that follow a pure rocky compositional Mass-Radius relation, as we discuss in Appendix D of Paper II.

\subsection{Envelope enrichment}
When the sublimation of the incoming solids is considered in core accretion, the amount of H-He that a core can bind changes \citep{HI11, Venturini15, Venturini16, Brouwers20}. 
In principle, for uniform envelope metallicity, the increase of mean molecular weight makes the  envelope more prone to contract, and therefore gas accretion is favoured \citep{Venturini16, Venturini17}.
However,  when Z is not assumed uniform, and the resulting compositional gradient is calculated self-consistently, the heat transport is modified. 
The interior becomes hotter, which abates slightly gas accretion 
\citep{Bodenheimer18}. Unlike water, silicates sublimate deeper in the envelope, so a non-uniform Z is expected.  
According to the results of \citet{Bodenheimer18} (the only who have managed to include compositional gradients in formation models so far), when silicate enrichment is considered, 
a core of $\Mcore$\footnote{With `core' we refer here to the mass of all metals.} $\approx 7 \, \Mearth$ should bind an envelope containing $\MHHe = 0.37$ \ME, 
compared to $\MHHe = 0.68 \Mearth$ when silicate enrichment is neglected and  all the solids are deposited in the core. 
In  other words, the effect of silicate enrichment seems to give H-He masses of only a factor 2 smaller. 
In the light of this results, silicate pebble enrichment is not expected to modify substantially the H-He masses that we find in this work.
Nevertheless, we consider that the inclusion of compositional gradients should be studied under much broader conditions to asses its true impact.

\section{Conclusions}
We studied pebble-driven planet formation inside the water ice line. 
For the first time, the pebble size and pebble surface density are computed self-consistently from dust growth and evolution calculations, 
instead of being prescribed as a free parameter as done by recent works \citep{Lambrechts19, Ogi20}.
An important result of this study is the finding that pebble accretion is extremely sensitive to disc turbulence within the water ice line. 
This is a direct consequence of dust growth being limited by fragmentation in this region of the disc. 
Turbulence promotes fragmentation, which makes the silicate pebbles attain very different sizes for different viscosities. 
This has dramatic consequences for the planet growth. For intermediate viscosities of $\alpha = 10^{-3}$, the initial moon-mass embryos practically do not grow at all.
We find that the typical output of silicate pebble accretion for low viscosities ($\alpha \lesssim 10^{-4}$) are terrestrial planets, ie. planets with a mass at the time of disc dispersal between that of Mars and $\sim$ 3 Earth masses. 
The formation of terrestrial planets by dry pebble accretion is very robust concerning disc metallicity and embryo location. 
Embryos growing in-situ from 0.1 au to 1 au, or migrating from just inside the water ice line, typically lead to terrestrial planets. 
In addition, discs with very low initial dust-to-gas ratio of Z = 0.005 (which would correspond to the lowest-metallicity stars in the solar neighbourhood), can also produce terrestrial planets, but less massive, of  $\sim$0.1-0.5 \ME.

Three factors limit the formation of purely rocky planets with masses exceeding  $\sim$5 \ME. 
First, the pebble isolation mass within the ice line decays typically fast to $\Miso \lesssim 5 \, \Mearth$ within the first million years of disc evolution. 
Second, after approximately 1 to 3 Myrs of disc evolution, photoevaporation carves a gap in the mid-disc. This truncates the  supply of pebbles from the outer disc, halting the core growth.  
Finally, even if the core manages to grow fast, for example in a very metallic disc and/or with extremely low viscosity, when $\Mcore \approx  5 \, \Mearth$, the protoplanet is able to bind a substantial H-He envelope, in some cases so massive that the planets are no longer core-dominated, but resemble more a sub-Saturn or gas-dwarf.  

Regarding the bi-modality of the Kepler size distribution, we find that all the rock-dominated planets that we form with mass below $\sim4\,\Mearth$ lose their atmospheres completely and contribute to the first peak of the size distribution. A second small group of planets, with cores between 4 and 5 \ME, end up with radius between 2 and 2.3 $\RE$. These planets lost typically $\sim$99\% in mass of their original envelopes. Remarkably, for planets with orbital periods below 100 days, we do not obtain any object with size between $\sim$2.5 and 6 $\RE$. Our results show that rocky planets can sometimes contribute to the valley and the lower part of the second peak, but do not seem to be able to account entirely for the second peak. We study the origin of the second peak in an accompanying letter (Paper II).

\bigskip
\small{
\textit{Acknowledgements}. We thank the anonymous referee for the criticism and enthusiasm about our work. We also thank B. Bitsch for his feedback and interest. J.V and O.M.G thank the ISSI Team "Ice giants: formation, evolution and link to exoplanets" for fruitful discussions. O.M.G acknowledges financial support from ISSI Bern in the framework of the Visiting Scientist Program. OMG is partially support by the PICT 2018-0934 and PICT 2016-0053 from ANPCyT, Argentina. OMG and MPR acknowledge financial support from the Iniciativa Cient\'{\i}fica Milenio (ICM) via the N\'ucleo Milenio de Formaci\'on Planetaria Grant. MPR also acknowledges financial support provided by FONDECYT grant 3190336. We thank J. Haldemann for technical help. This work has in part been carried out within the framework of the NCCR PlanetS supported by the Swiss National Science Foundation. 
}

% ======================================================================================================================================

\bibliographystyle{aa}
\bibliography{lit_2020}

\end{document}